\documentclass[jgr]{agutex}
\usepackage{graphicx}
\usepackage{amsmath}
\usepackage{color}
\usepackage[usenames,dvipsnames]{xcolor}

\authorrunninghead{Lamy et al.}
\titlerunninghead{Multi-spectral diagnosis of Saturn's aurorae}

\begin{document}

\title{Multi-spectral simultaneous diagnosis of Saturn's aurorae throughout a planetary rotation}

\authors{L. Lamy\altaffilmark{1}, R. Prang\'e\altaffilmark{1}, W. Pryor\altaffilmark{2}, J. Gustin\altaffilmark{3}, S.~V. Badman\altaffilmark{4}, H. Melin\altaffilmark{5}, T. Stallard\altaffilmark{5}, D.~G. Mitchell\altaffilmark{6},  P.~C. Brandt\altaffilmark{6}}

\altaffiltext{1}{LESIA-Observatoire de Paris, CNRS, Universit\'e Pierre et Marie Curie, Universit\'e Paris Diderot, 5 Place Jules Janssen, 92195 Meudon, France}
\altaffiltext{2}{Central Arizona College, Coolidge, Arizona 85128, USA}
\altaffiltext{3}{Laboratoire de Physique Atmosph\'erique et Plan\'etaire, Universit\'e de Li\`ege, Li\`ege, B-4000, Belgium}
\altaffiltext{4}{Institute of Space and Astronautical Science, JAXA, Sagamihara, Japan}
\altaffiltext{5}{Department of Physics and Astronomy, University of Leicester, Leicester LE1 7RH, UK}
\altaffiltext{6}{Johns Hopkins University Applied Physics Laboratory, Laurel, MD 21042, USA}

\begin{abstract}
From the 27th to the 28th January 2009, the Cassini spacecraft remotely acquired combined observations of Saturn's southern aurorae at radio, ultraviolet and infrared wavelengths, while monitoring ion injections in the middle magnetosphere from energetic neutral atoms. Simultaneous measurements included the sampling of a full planetary rotation, a relevant timescale to investigate auroral emissions driven by processes internal to the magnetosphere. In addition, this interval coincidently matched a powerful substorm-like event in the magnetotail, which induced an overall dawnside intensification of the magnetospheric and auroral activity. We comparatively analyze this unique set of measurements to reach a comprehensive view of kronian auroral processes over the investigated timescale. We identify three source regions in atmospheric aurorae, including a main oval associated with the bulk of Saturn Kilometric Radiation (SKR), together with polar and equatorward emissions. These observations reveal the co-existence of corotational and sub-corototational dynamics of emissions associated with the main auroral oval. Precisely, we show that the atmospheric main oval hosts short-lived sub-corotating isolated features together with a bright, longitudinally extended, corotating region locked at the southern SKR phase. We assign the susbtorm-like event to a regular, internally driven, nightside ion injection possibly triggerred by a plasmoid ejection. We also investigate the total auroral energy budget, from the power input to the atmosphere, characterized by precipitating electrons up to 20~keV, to its dissipation through the various radiating processes. Finally, through simulations, we confirm the search-light nature of the SKR rotational modulation and we show that SKR arcs relate to isolated auroral spots. We characterize which radio sources are visible from the spacecraft with the fraction of visible southern power estimated to a few percent. The resulting findings are discussed in the frame of pending questions as the persistence of a corotating field-aligned current system within a sub-corotating magnetospheric cold plasma, the occurrence of plasmoid activity and the comparison of auroral fluxes radiated at different wavelengths.
\end{abstract}

\begin{article}

\section{Introduction}

Saturn's aurorae have been increasingly studied in the past decade with the arrival of the Cassini spacecraft in orbit. These powerful electromagnetic emissions, radiated from high-latitude regions surrounding the magnetic poles, are analogous to those observed at Earth and at the other giant planets \citep[and refs therein]{Bhardwaj_RG_00}. They are generated by accelerated charged particles flowing along auroral magnetic field lines, mainly electrons ranging from a few to a few tens of keV, and are made up of radiations within the UltraViolet (UV), visible and InfraRed (IR) domains, initially originating from collisional processes between precipitating particles and the upper atmosphere, and Saturn's Kilometric Radiation (SKR) at radio wavelengths, radiated through wave-electron cyclotron resonance up to a few Saturn radii above the atmosphere (1~R$_S$~=~60268~km) \citep[and refs therein]{Kurth_09}. These emissions have often been studied separately, providing complementary insights on auroral processes, while leaving open questions on their relationship.



UV (and visible) radiations reveal the instantaneous response of the kronian neutral atmosphere, dominated by atomic and molecular hydrogen, to auroral precipitation. Their emission rate is directly proportional to the incident electron flux. In the 80 to 170~nm range, they mostly consist of the H Lyman-$\alpha$ (Ly-$\alpha$) transition at 121.6~nm and H$_2$ continuum, Lyman and Werner bands. These emissions are primarily organized along southern and northern conjugate main auroral ovals, whose origin is attributed to field-aligned currents driven by the shear between open and closed field lines \citep{Cowley_AG_04,Bunce_JGR_08}. On average, they are located around $-72^\circ$ kronocentric latitude in the south, circularly-shaped under quiet conditions, and brighter/narrower on the dawn sector. They are fed by precipitating electrons up to 20~keV and radiate 1-100~kR in the H$_2$ bands \citep{Trauger_JGR_98,Gerard_JGR_04,Badman_AG_06,Gustin_Icarus_09,Lamy_JGR_09,Nichols_GRL_09}. Their dynamics are primarily controlled by the solar wind and the fast planetary rotation. On the one hand, interplanetary forward shocks can trigger sudden brightenings of the midnight-to-dawn polar cap and give rise to spiral-shaped structures \citep{Prange_Nature_04,Clarke_Nature_05,Clarke_JGR_09}. On the other hand, the intensity distribution and the position of north/south main ovals are modulated at the corresponding hemispheric SKR period \citep{Sandel_Science_82,Nichols_GRL_10,Nichols_GRL_10b}, while isolated features are at times observed to sub-corotate at velocities ranging from 30 to 80\% of the full rotation rate \citep{Grodent_JGR_05,Gerard_JGR_06,Grodent_JGR_11}. Beyond this main source region, equatorward emissions include a 1-2~kR bright secondary ring of emission at $-67\pm3.5^\circ$, suspected to be fed in by suprathermal electrons populating the inner magnetosphere \citep{Grodent_JGR_10}, and an intermittent spot at the footprint at the Enceladus flux tube resulting from the planet-satellite electrodynamic coupling \citep{Pryor_Nature_11}. Sporadic and fainter emissions have also been recorded poleward. Among them, some arcs of emission connecting to the main oval (so-called bifurcations) have been interpreted as the signature of transient reconnection events at the magnetopause \citep{Radioti_JGR_11}. Overall, UV aurorae radiate a few 10$^{10}$ W in each hemisphere, corresponding to a power input approximately one order of magnitude larger \citep[and refs therein]{Clarke_JGR_09}.

Aurorae observed in the near IR are radiated in the ionosphere by quasi-thermalized H$_3^+$ ions, produced from the ionization of molecular hydrogen by electron impact. Their intensity is proportional to the square root of the incident electron flux and is strongly dependent on the temperature \citep{Tao_Icarus_11}. H$_3^+$ ro-vibrational lines lie between 1 and 5~$\mu$m \citep{Neale_ApJ_96}. Their observation yielded estimates of thermospheric temperatures around 400-500K and H$_3^+$ collisional excitation time and lifetime of 26 and 500s respectively \citep{Melin_Icarus_07,Melin_GRL_11}. H$_3^+$ aurorae also organize along a main oval, colocated to the UV one, similarly enhanced at dawn and modulated at the corresponding hemispheric SKR period, which locally radiates 0.4~mW.m$^{-2}$ \citep{Stallard_ApJ_10,Badman_GRL_11,Melin_GRL_11,Badman_JGR_12b}. A secondary oval has been detected at lower latitudes with intensities four times less than the main oval, possibly conjugate to the UV one, but here proposed to result from field-aligned currents driven by the plasma co-rotation breakdown at 3-4~R$_S$ \citep{Stallard_Nature_08a}. Variable emissions within the polar cap (isolated spots/arcs or diffuse emission) with intensities similar to those of the main oval have been interpreted as a feature unique to H$_3^+$ \citep{Stallard_Nature_08b}. A first estimate of the overall power radiated by planetwide H$_3^+$ emission yielded a few 10$^{11}$~W \citep{Stallard_ApJ_99}.

Above the atmosphere, SKR is driven by the Cyclotron Maser Instability, which amplifies radio waves close to the local electron cyclotron frequency (which is proportional to the magnetic field strength) between a few kHz and 1000~kHz to large amplitudes from resonant electrons. They are radiated at large angles from the local magnetic field \citep{Wu_ApJ_79,Kaiser_Science_80}. 
The bulk of SKR emission corresponds to X-mode radio sources populating field lines connected to the main oval, with a strong Local Time (LT) dependence of the source intensity, peaking around 08$:$00~LT. The total radiated power reaches a few $10^7$~W.sr$^{-1}$ on average \citep{Lamy_JGR_08a,Lamy_JGR_09}. In situ measurements recently acquired within low frequency SKR sources identified downward beams of 6-9~keV electrons as the source of the emission with an electron-to-wave conversion efficiency of $\sim$1\% \citep{Lamy_GRL_10,Mutel_GRL_10,Lamy_JGR_11,Kurth_PRE7_11,Schippers_JGR_11,Menietti_JGR_11}. SKR intensity is strongly controlled by the solar wind \citep{Desch_JGR_83} and it is rotationally modulated at a period proper to each hemisphere through the rotation of an active region \citep[and refs therein]{Desch_GRL_81,Gurnett_GRL_09,Lamy_PRE7_11}. Superimposed to this rotational modulation, arc-shaped structures in dynamic spectra were found to result from localized active magnetic field lines in sub-corotation at a $\sim$90\% level \citep{Lamy_JGR_08b}.

Energetic Neutral Atoms (ENA), essentially hydrogen and oxygen, are produced when energetic ions undergo charge-exchange collisions with cold neutral particles. In the middle nightside magnetosphere, intensifications of ENA mapping to the ring current indicate recurrent injections of energetic ions. These emissions are anisotropically emitted, primarily along the direction of incident ions (Compton-Getting effect \citep{Compton_PR_35}). ENA are known to intensify quasi-periodically on the nightside region and then co-rotate with the planet at the southern (hereafter S) SKR period \citep{Krimigis_Science_05,Paranicas_GRL_05,Carbary_GRL_08,Carbary_GRL_11a}. Investigating the rise of a bright ENA event magnetically connected to a co-rotating UV bright region, \citet{Mitchell_PSS_09} suggested that ENA intensifications result from quasi-periodic injections near local midnight, driven by current sheet reconnection and subsequent plasmoid release down the tail.\citet{Brandt_GRL_10} proposed that field-aligned currents driving aurorae result from pressure-driven gradients forming a partial ring current, whose strength could be diagnosed (through plasma pressures) by ENA measurements.

In this study, we investigate two days of combined observations of Saturn's aurorae at radio, UV and IR wavelengths, and their relationship with a reservoir of energetic particles in the middle magnetosphere. Within this time period, simultaneous measurements sampled a full planetary rotation, a relevant timescale to investigate auroral emissions driven by processes internal to the magnetosphere. This interval coincidently matched a powerful susbtorm-like event in the magnetotail, which induced an overall intensification of magnetospheric and auroral activity. We present a comparative analysis of this unique set of complementary measurements to reach a comprehensive view of kronian auroral processes, in terms of source regions (section \ref{sources}), dynamics (section \ref{dynamics}), energy transport and dissipation (section \ref{energy}), and SKR modeling (section \ref{skr_modeling}). These results are then discussed in section \ref{discussion}.

\begin{figure*}
\centering\includegraphics[width=1.1\textwidth,angle=-90]{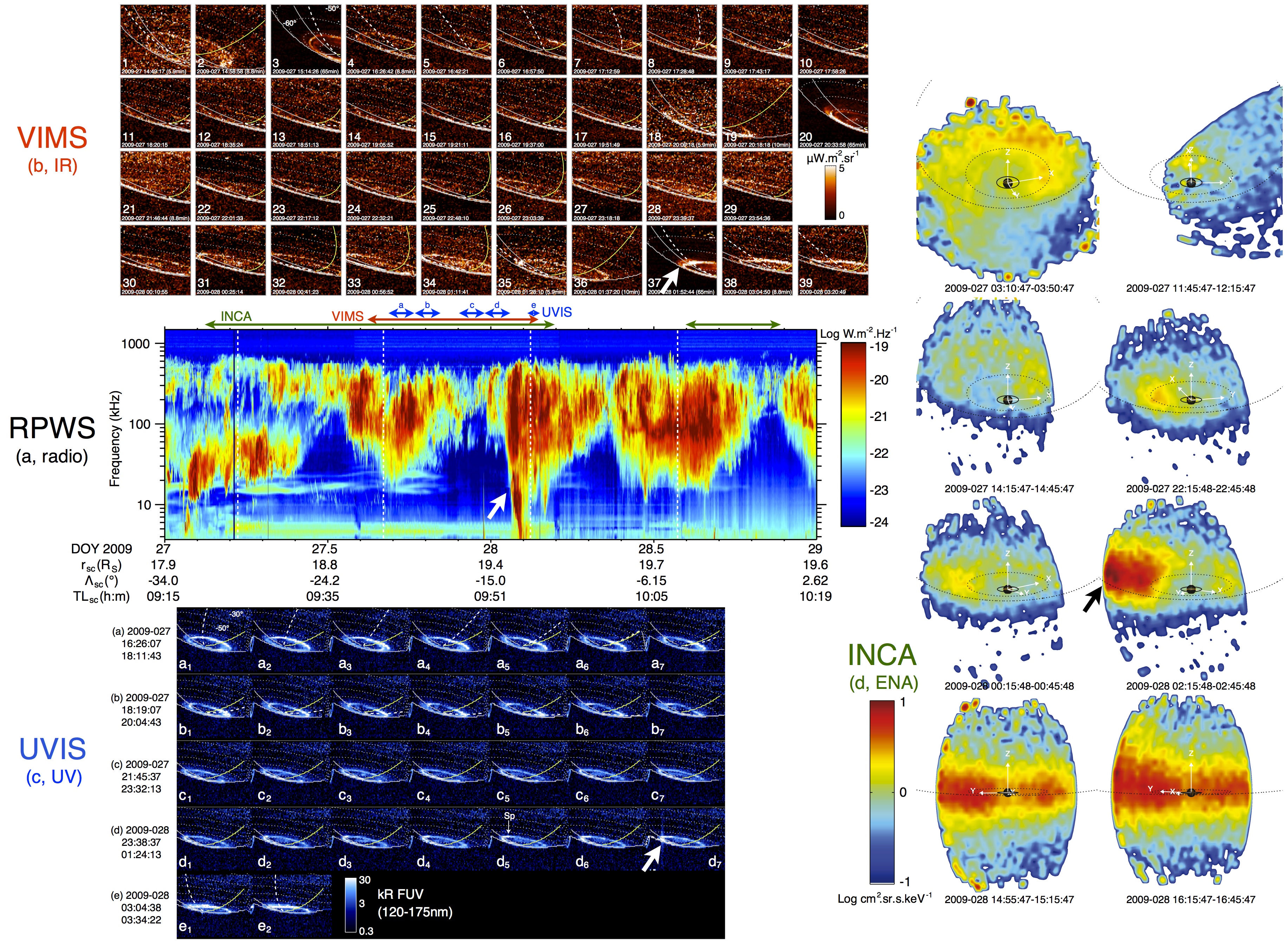}
\caption{Summary of Cassini observations from day 2009-027 to 2009-028. (a) RPWS dynamic spectrum, dominated by SKR emission between 3.5 and 800~kHz. Vertical dashed lines mark the maxima predicted by the S SKR phase system. Colored double arrows indicate simultaneous observations of VIMS (red), UVIS (blue) and INCA (green) instruments. (b) Sequence of H$_3^+$ images acquired with VIMS. Intensities were integrated over 5 spectral channels (3.417, 3.450, 3.532,  3.612 and 3.667~$\mu$m) dominated by the auroral signal and processed as described in the main text, with the exception of the correction for limb brightening. (c) Five sequences (labelled a to e) of H Ly-$\alpha$ and H$_2$ images obtained with UVIS. Intensities have been processed as described in the main text, integrated from 120 to 190~nm and plotted with a logarithmic scale. Sp indicates an isolated UV hot spot. Kronocentric grids of coordinates, derived at the altitude of peak aurorae, have been superimposed to images. Gray (yellow respectively) solid lines mark midnight, dawn and dusk (noon respectively) meridians, and dotted lines mark latitudes every 10$^\circ$. The dashed line indicates the rotating guide meridian built from the S SKR phase system (see text). (d) Snapshots of ENA images of H$^+$ ($24-55$~keV) acquired by INCA. The white rotating frame is built with Z along the rotation axis and Y along the rotating guide meridian, X completing the right-handed frame. The dotted lines give the orbits of Dione and Titan. Large arrows indicate a simultaneous powerful auroral and dawnside ENA brightening at the beginning of DOY 28.}
\label{fig1}
\end{figure*}

\section{Cassini observations}
\label{observations}

This section briefly describes the set of Cassini observations employed, including radio, UV, and IR aurorae as well as measurements of Energetic Neutral Atom (ENA). 

\subsection{Ephemeris}
\label{ephemeris}

During the 27-28 January 2009 time interval, hereafter expressed in days of year (DOY) 2009, the Cassini spacecraft passed from orbital revolution (Rev) 101 to 102, reaching apokrone at DOY 28 17$:$12~UT.  Its distance to the planet's center $r_{sc}$ remained between 17.9 R$_S$ and 19.7 R$_S$ at apokrone, while its local time LT$_{sc}$ varied from 09$:$15 to 10$:$19 and latitude $\Lambda_{sc}$ from $-34^\circ$ to $+2.6^\circ$. The spacecraft was therefore able to record Saturn's southern (hereafter S) aurorae from an approximately fixed location, only varying in latitude.

\subsection{RPWS}

Saturn's low frequency radio emissions are continuously monitored by the Radio and Plasma Wave Science (RPWS) experiment onboard Cassini \citep{Gurnett_SSR_04}. Its High Frequency Receiver (HFR) scans frequencies between 3.5~kHz and 16~MHz, which includes the SKR spectral range. Here, RPWS/HFR operated with a spectral resolution $\Delta f/f$~=~5~\%, integration times $\Delta t$~= 0.25 to 1~s and logarithmically-spaced frequency channels from 3.5 to 320~kHz, and $\Delta f/f~\sim$~2 to 8\%, $\Delta t$~=~0.08~s and linearly spaced channels from 320 to 800~kHz. The corresponding temporal resolution, given by the delay between two successive HFR scans at a given frequency, varied between 16 and 32~s along the interval. Continuous observations hereafter refer to uninterrupted temporal sampling down to these time scales. 

This combination of good spectral and temporal resolution served to build the detailed time-frequency pattern displayed in Figure \ref{fig1}a, where the intensity is the Poynting flux S, derived from a standard 2-antenna goniopolarimetric inversion \citep{Cecconi_RS_05}, and normalized to 1 astronomical unit \citep{Lamy_JGR_08a}.



\subsection{UVIS}

The Cassini UltraViolet Imaging Spectrograph (UVIS) \citep{Esposito_SSR_04} is capable of observing Saturn's upper atmosphere between 112 and 191~nm with its Far-UltraViolet (FUV) channel, through a slit composed of 64 spatial pixels. From DOY 27 16$:$25$:$41 to DOY 28 03$:$04$:$12~UT, UVIS was mainly used in the low spectral resolution slit configuration, with a spectral resolution $\lambda/\Delta\lambda\sim$~60 and 32 spectral resolution elements, rebinned onboard from 1024 original spectral elements. Each of the 64 spatial pixels had a 1~mrad field-of-view (FOV) along the slit and 1.5~mrad across it. 

The observations were split into five successive sequences (lines labelled a to e in Figure \ref{fig1}c), with the slit (y direction) being approximately oriented in the south-north direction. Sequences a to d (e respectively) comprise 793 (224 respectively) consecutive 8~s long exposures, slewing Saturn's S pole in the direction perpendicular to the slit. West-to-east slow slews (left-to-right in Figure \ref{fig1}c) provided 30 complete maps of the auroral region (numbered from 1 up to 7 for each sequence), while interspersed east-to-west quicker slews provided 25 additional snapshots (unlabelled). Pixels along the slit (y axis) provide simultaneous measurements, and a full scan of the auroral oval required approximately 11~min for west-east slews. Maps a$_1$ to e$_2$ eventually provide a set of spectro-images (2D maps with spectral information for each pixel) integrated over 8~s, but acquired over a duration of $\sim$11~min, that form the basis of our analysis. Within a sequence, successive images were separated by 14~min, taken as the typical temporal resolution.

This dataset was first calibrated through the pipeline of the UVIS team, including corrections for the flat-field response and the instrumental background. The bandwidth of the UVIS FUV channel includes auroral emissions (mostly the H Ly-$\alpha$ line at 121.6~nm, the Lyman and Werner bands and the Lyman continuum of H$_2$) which can be subject to absorption by overlying hydrocarbons layers, together with a non negligible background. To retrieve the auroral signal we are interested in, we applied the supplementary data processing described and discussed in appendix \ref{uvis_processing}. Figure \ref{fig1}c displays processed maps integrated over the 120 to 175~nm spectral range. The kronocentric coordinates of spatial pixels were derived from spice kernels at an altitude of 1100~km, where the UV aurorae usually peak \citep{Trauger_JGR_98,Lamy_these_08,Gerard_GRL_09}. 

\subsection{VIMS}

The Visual and Infrared Mapping Spectrometer (VIMS) \citep{Brown_SSR_04} is a one spatial pixel spectrometer, with a 0.5x0.5~mrad FOV, covering the spectral range 0.8 to 5.1~$\mu$m with 256 spectral pixels. VIMS continuously observed part of the S auroral region between DOY 27 14$:$49$:$17 and DOY 28 03$:$29$:$37~UT, with a spectral resolution of $\lambda/\Delta\lambda\sim$~200. These observations were organized as 39 successive spectro-images of 64x64 spatial pixels (numbered from 1 to 39 in Figure \ref{fig1}b), corresponding to a total FOV of 32x32~mrad. 

Individual images were acquired with integration times between 5.9 and 65~min, corresponding to integration times of $\sim$0.09 to $\sim$0.9~s per pixel, with the observing time increasing from bottom to top and from left to right. They were separated by varying intervals, reaching $\sim$15~min on average.

The original VIMS dataset was calibrated with the pipeline developped at the University of Arizona (replicated on the NASA PDS archive). The emission of H$_3^+$ was obtained by summing up the intensity over 5 wavelength bins (3.417, 3.450, 3.532,  3.612 and 3.667~$\mu$m), which include auroral H$_3^+$ together with some significant background. Similarly to UVIS, we applied a specific data processing, described in appendix \ref{vims_processing} to isolate the emission radiated by auroral $H_3^+$ only, as illustrated in Figure \ref{fig1}b, and then corrected for limb brightening. The kronocentric coordinates of spatial pixels were derived from NASA NAIF spice kernels at 1150~km, to match the H$_3^+$ peak altitudeÊ\citep{Stallard_GRL_12}.


\subsection{INCA}

The Ion and Neutral CAmera (INCA) of the Magnetospheric IMaging Instrument (MIMI) \citep{Krimigis_SSR_04} obtains remote ENA images of the H$^+$ ($\sim$~7-300~keV) and O$^+$ ($\sim$~50-400~keV) ion distributions over a FOV of 90$\times$120$^\circ$. The effective angular resolution is about 8$\times$4$^\circ$ at 50~keV H. In this paper we analyze hydrogen ENA images obtained in the 24-55~keV range, which can be considered as an optimal energy range in terms of intensity and angular resolution. ENA images were integrated over 15~min, implying a smearing in LT of about 6$^\circ$ caused by the corotational drift of the ion distribution. Figure \ref{fig1}d shows a sequence of ENA images obtained during DOY 27 and 28.



\section{Observed auroral source regions}
\label{sources}

Original data shown in the summary Figure \ref{fig1} reveal a variety of auroral source regions and dynamics. To investigate them in further details, we built the supplementary animation S$_1$, which displays the overlapping radio, UV and IR observations  (including strictly simultaneous ones). For each frame, the UVIS and VIMS observing time intervals are indicated by blue and red boxes on the RPWS dynamic spectrum. The UV and IR data have been processed using the pipeline described in appendices \ref{uvis_processing},\ref{vims_processing} and separated in physical species (H and H$_2$ for UV, and H$_3^+$ for IR). They are represented both in the instrumental FOV (top), and as polar projected views (bottom). 

The main oval is the brightest source region, known to be connected with the bulk of SKR emission. It remains roughly circular at latitudes down to $-70^\circ$, indicating quiet conditions, at the exception of the LT quadrant lagging behind the brightest region, which is localized a few degrees poleward. This fits previous observations of the oscillating motion of the auroral oval, sunward-shifted at SKR maxima \citep{Provan_JGR_09b,Nichols_GRL_10b}. There is a clear spatial colocation of H, H$_2$ and H$_3^+$ radiations at timescales down to $\sim$12-15~min, over which most of the images were built. This result is strengthened by the comparison of successive strictly simultaneous UVIS and VIMS pixel measurements along the 'swath of simultaneity' (not shown), which indicate clearly colocated main ovals, as found previously \citep{Melin_GRL_11}. With respect to the latter study, H Ly-$\alpha$ images are here noisier but more reliable since they have been corrected for any H$_2$ contamination of the Ly-$\alpha$ line. The dynamics of the oval are mainly modulated by the planetary rotation, as investigated further below. It is composed of a bright extended region and fainter emission along the rest of the oval, including isolated features of variable intensity.

Polar emissions form a second bright emitting region. Interestingly, they are not only observed in H$_3^+$ emissions (frames 1-3, 20-35 in animation S$_1$), whose intensities compare to those of the main oval, but also for H and H$_2$, with roughly similar and weaker relative intensities respectively (see frames 26-30). When observed together (frames 21-34), these radiations are still spatially coincident. They are seen on the dayside hemisphere, peaking close to noon. The observed polar features (spots, arcs) remain organized along a longitudinally extended ring of emission between $-75^\circ$ and $-82^\circ$ latitude and from 06$:$00 to 16$:$00~LT. This ring seems to sometimes connect to the main oval (frames 2-3, 23 or 27) along a spiral structure. For the last two examples, the spiral has formed posterior to a discontinuity of the main oval in the noon-to-dusk quadrant (frames 17-25). Such polar emissions may be associated with lobe reconnection at high latitude, sporadically extending to the main oval by merging to the region of closed flux \citep{Badman_JGR_12a}.


Finally, a fainter, diffuse, ring of emission appears equatorward of the main oval in H and H$_2$ images. It is best visible on images of sequences a and b, for which $\Lambda_{sc}$ remained close to $-20^\circ$ : a dawnside arc is visible at latitudes of $\sim$~$-70^\circ$ (in images a$_1-$b$_7$), while a duskside diffuse arc appears at latitudes down to $-65^\circ$ (a$_5-$a$_6$ or b$_1$, more contrasted at H Ly-$\alpha$). These arcs, separate from the main emission, may represent different segments of a common low-latitude ring, whose intensity drops through noon. The dawnside arc could alternately be the end of a spiral-shaped main oval, but it does not appear to corotate nor change in intensity with time. In any case, this reveals that the secondary oval previously identified on the nightside at $-67\pm3.5^\circ$ \citep{Grodent_JGR_10} or $\sim$~$-70^\circ$ \citep{Melin_GRL_11} extends toward the dayside hemisphere. Because of the low signal-to-noise ratios (SNR) of IR observations corresponding to UV sequences a and b, it was not possible to directly identify a simultaneous H$_3^+$ counterpart. However, all three H$_3^+$ images with long integration times (3, 20, 39) clearly show such an equatorward arc at $-70\pm1^\circ$ (from 00$:$00 to 08$:$00~LT, coincident with the dawnside arc of UV images a$_1$), $-68\pm1^\circ$ (from 03$:$00 to 09$:$00~LT) and $-70\pm1^\circ$ (from 03$:$00 to 07$:$00~LT) respectively. This strengthens the previous identification of equatorward aurorae from ground-based IR observations on the dawn and dusk sides \citep{Stallard_Nature_08a}.

No Enceladus-related emissions could be identified during this interval.


\section{Dynamics}
\label{dynamics}

\begin{figure*}
\centering\includegraphics[width=1.\textwidth]{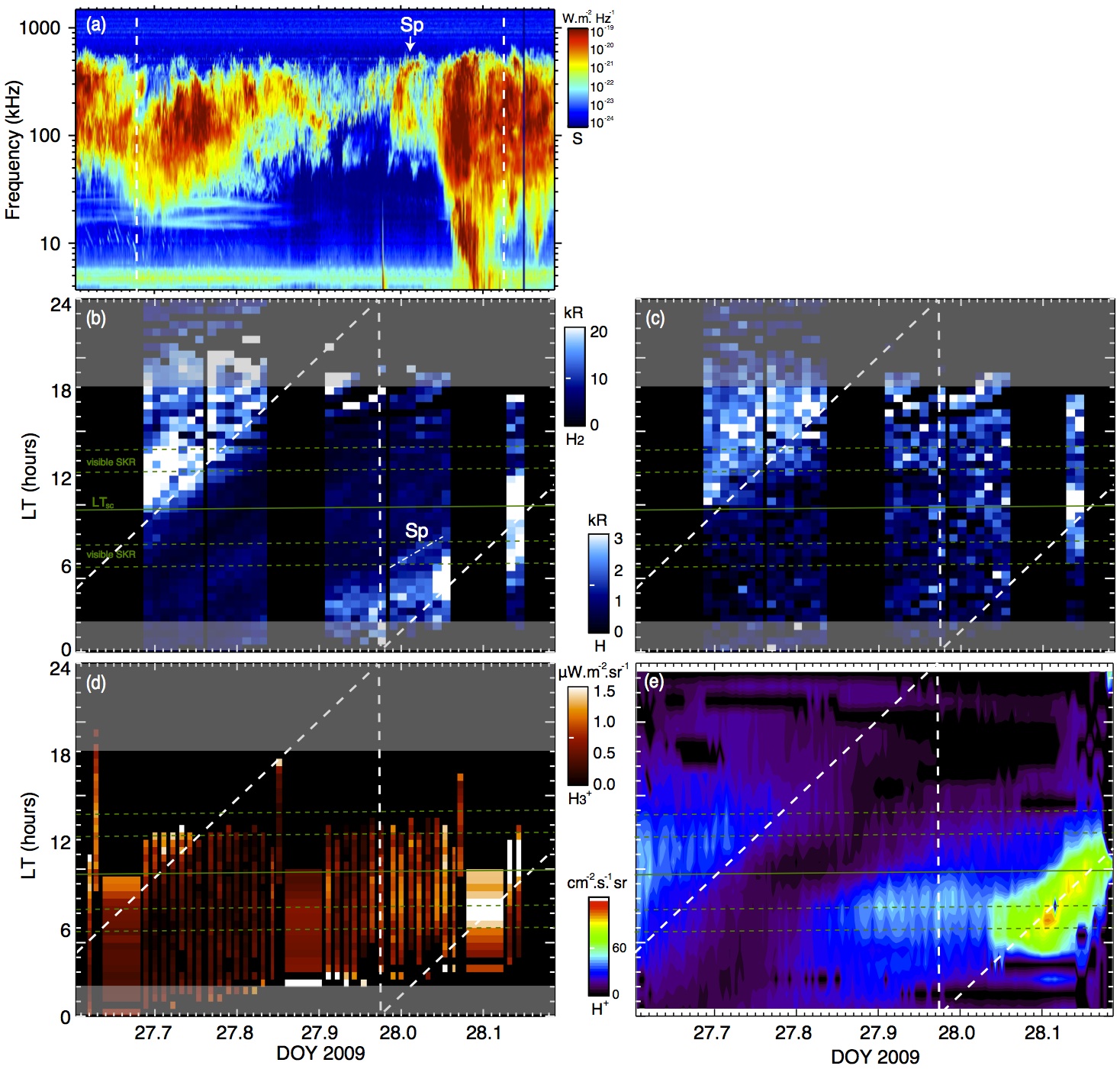}
\caption{(a) RPWS dynamic spectrum (identical to Figure \ref{fig1}a) restricted to UVIS/VIMS observing intervals, with vertical dashed lines marking the predicted S SKR maxima. (b) Total brightness of H$_2$ aurorae plotted in the time-LT frame of coordinates. The emission has been corrected for hydrocarbon absorption (see appendix \ref{uvis_processing}) and averaged over the main oval, namely between $-77^\circ$ and $-65^\circ$ latitude. The dashed line marks the rotating reference meridian derived from the S SKR phase.  Gray shaded regions indicate LT ranges corresponding to limb emissions. The solid green line indicates LT$_{sc}$, the sub-spacecraft LT. The two pairs of dashed green lines delineate the two LT ranges where SKR sources are expected to be visible, on both sides of LT$_{sc}$, as described in section \ref{skr_modeling}. Sp indicates a UV hot spot. Its motion, fitted by the white dashed-dotted line, through the bottom pair of green lines coincides with a powerful vertex-early SKR arc, also labelled S. (c) Same as (b), but for the H Ly-$\alpha$ brightness, also corrected for hydrocarbon absorption. (d) Same as (b), but for the H$_3^+$ flux (processed as described in appendix \ref{vims_processing}). (e) Observations of H$^+$ ($24-55$~keV) at highest time resolution, projected along the line-of-sight onto the equator, corrected for the Compton-getting effect \citep{Carbary_JGR_08} and averaged over radial distance and LT.}
\label{fig2}
\end{figure*}

Continuous observations of SKR monitor the auroral context at high temporal resolution, revealing variations at different timescales.

\subsection{Rotational modulation}

The kilometric radiation is firstly dominated by the rotational modulation, as evidenced by the presence of four consecutive bursts every $\sim~11$~hours. The prevailing left-handed circular polarization of radio waves (not shown) indicates extraordinary mode emission originating from the S hemisphere, as expected from visibility conditions at negative latitudes. The SKR maxima occur regularly, close to the predictions of the S SKR phase system (vertical dashed lines) although this system provides a long-term guide built from 200-days averaged data \citep{Lamy_PRE7_11} and thus neglects possible phase variations at shorter timescales. The same study showed that the S SKR rotational modulation is produced by a search-light active region rotating at the S SKR period, which is strongly LT dependent, as the intensity of dawnside radio sources can increase by several orders of magnitude. The S SKR maxima thus generally reflect the transit through dawn of an active sector, $\ge90^\circ$ extended in longitude. Under the assumption that SKR bursts occur when the active rotating region precisely reaches 08~LT, where the brightest SKR sources reside \citep{Lamy_JGR_09}, the S SKR phase system directly translates into a reference rotating meridian.

This guide meridian is superimposed onto UV and IR images of Figures \ref{fig1}b,c and onto the frames of supplementary animation S$_1$ as white dashed lines. The main oval clearly displays a bright region which always lies close to the rotating SKR phase meridian. This is clearer for UV observations, which fully sample the auroral region, and in particular in sequences a, b or e, for which the activated portion of the oval is best visible from Cassini. With respect to previous work ($e.g.$ \citep{Nichols_GRL_10}), the UV modulation here displays a larger amplitude associated with a well-defined active region. A similar rotating region appears in the H$_3^+$ main oval, as in image 37. It is less straightforward (lower SNR, incomplete FOV), but clearly supported by a comparable morphology of the main oval in the combined UV observations (frames 6-10 and 34-39 in animation S$_1$). This is in line with the recent identification of the H$_3^+$ rotational modulation in a dedicated study \citep{Badman_JGR_12b}. 

The search-light modulation effect can be investigated more quantitatively with Figure \ref{fig2}, which plots the intensity of H, H$_2$ and H$_3^+$ main ovals as a function of LT and time. In the three spectral ranges, but especially in the H$_2$ bands, the LT of the active region (saturated white pixels in Figures \ref{fig2}b,c,d) varies linearly with time throughout the time period, staying roughly parallel to the reference rotating meridian (white dashed line). The active region thus rotates at the S SKR period, while peaking $30^\circ$ in longitude (2 hours LT) ahead of the S SKR phase meridian with a $50^\circ$ to $90^\circ$ width. Its intensity is much larger than the average intensity for the rest of the oval whatever the considered species, and again best marked for H$_2$. This may come from either more energetic (typically $\ge10$~keV) electrons and/or larger precipitating fluxes. This point is discussed further in section \ref{energy}. Meanwhile, these results agree with the picture of a longitudinally extended upward field-aligned current (FAC) rotating at the S SKR phase, which would activate a localized portion of the auroral oval \citep{Andrews_JGR_10a}.

Interestingly, a $\sim$~$40^\circ$ wide region of the main oval of weaker intensity, best observed in Figure \ref{fig2}b, lags behind the active region by $\sim$~$90^\circ$. Such a "dark" region is consistent with downward FAC, lagging the upward FAC region by $\sim$~$90^\circ$, as observed in situ by \citet{Andrews_JGR_10a}.



Finally, INCA measurements displayed in Figure \ref{fig1}d and animation S$_1$ show enhanced ENA emissions variable with time within the ring current. Around the first SKR burst of DOY 27, intense ENA drifting through the pre-noon sector appear from 02$:$30~UT, indicating an anterior, large-scale, injection of energetic ions. This ion distribution has drifted to the nightside at around 08$:$00~UT but the spacecraft attitude changed such that most of the ring current fell outside of the FOV. However, at around 10$:$00~UT, an ion distribution of moderate ENA intensity could be seen drifting again through the pre-noon sector. From other observations and considering charge-exchange lifetimes \citep{Brandt_GRL_08}, it is most likely related to a new injection. Following the second SKR burst, at around 20$:$00~UT, a gradual ENA brightening occurred and remained localized post midnight until the end of the day. This non-rotating feature is visible until a dramatic intensification in the dawn sector at about 23$:$30~UT, which subsequently drifts through noon. This massive injection is well correlated both in intensity and time with the third, strongest, SKR burst of the DOY 27-28 interval. Following a 10~h long data gap on DOY 28, we finally note the presence of an ENA-emitting region drifting on the duskside around 14$:$30~UT, indicating another anterior, large-scale injection in the midnight-to-noon sector, in agreement with the fourth SKR burst. The phase relationship between ENA and auroral emissions can be quantitatively investigated with the equatorial projection of INCA observations displayed in Figure \ref{fig2}e. Due to low observing latitudes and vertically extended ion distributions, the retrieved LT have a low accuracy which decreases with time. Nonetheless, Figure \ref{fig2}e clearly shows the two ENA components (respectively corotating and fixed in LT) noticed above. The steady-state component, fixed at 07$:$30~LT and extending between 03$:$00 and 12$:$00~LT at the end of DOY 27, may be associated with auroral emissions observed along the main IR oval at similar times and LT in Figure \ref{fig2}d. The bright rotating component appears strikingly close to the S SKR phase (white dashed line), which makes the center of the auroral active region coinciding with the center of the rotating energetic ion distribution. The marked ENA brightening associated with the third SKR burst is further discussed below. 



\subsection{Plasmoid activity?}
\label{plasmoid}

We first note that the SKR bursts observed during DOY 26 and 29, preceding and following those investigated in this study, display lower intensities. Throughout our interval, the four well defined SKR bursts were all associated with energetic ions rotating through dawn. However, they have different characteristics. During the second and fourth bursts, the SKR spectrum was typical, extending from 20 to 800~kHz with peak intensities between 100 and 400~kHz. During the first burst, emission at high frequencies was missing and the spectrum rather peaks at lower frequencies, between 10 and 70~kHz. This point is discussed in section \ref{skr_modeling}. 

\begin{figure*}
\centering\includegraphics[width=1.\textwidth]{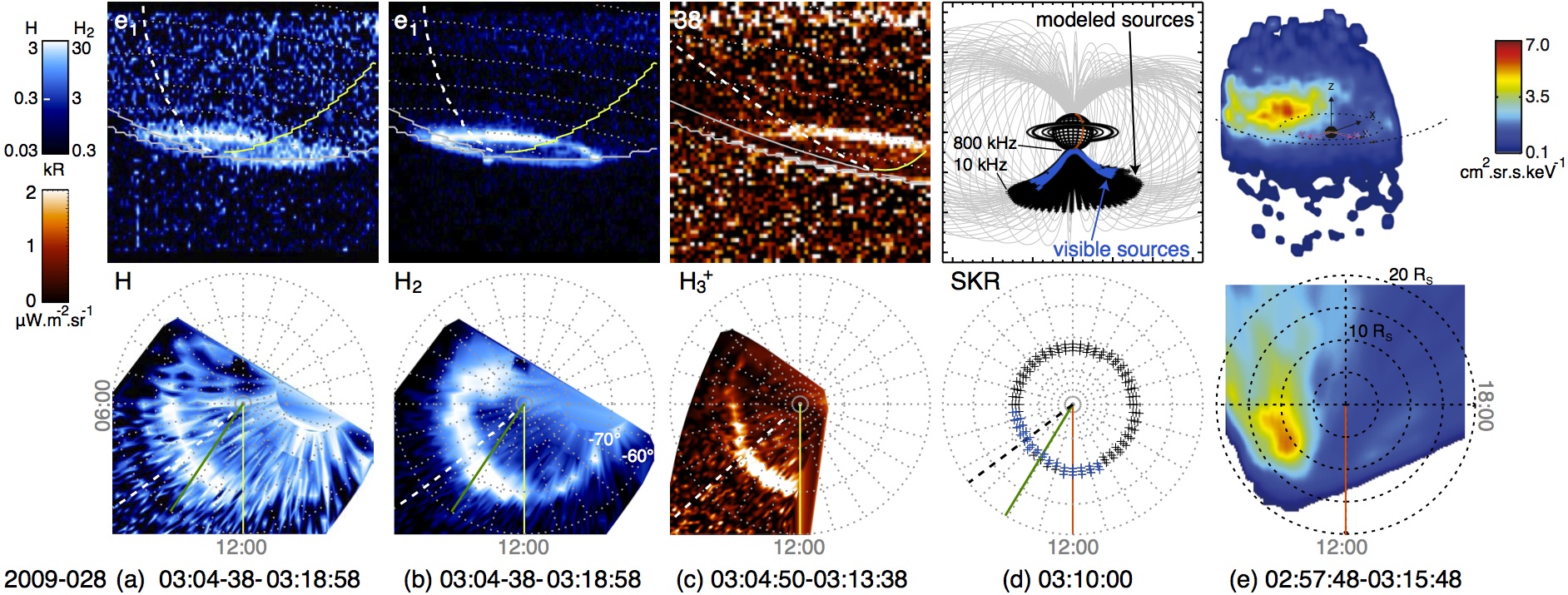}
\caption{Simultaneous observations during the rising phase of the powerful nightside ion injection at the beginning of DOY 28. (a,b,c) UV H-Ly-$\alpha$ and H$_2$ and IR H$_3^+$ aurorae represented in UVIS and VIMS FOV (top) and polar projected (bottom). These panels are identical to those of animation S$_1$ (frame 34). (d) SKR sources at 01$:$20~UT, as simulated by assuming the UV oval e$_1$ as the source region and parameters detailed in section \ref{skr_modeling}. Only a fraction of radio sources (black crosses) is visible from the spacecraft (blue crosses). (e) ENA images of H$^+$ ($24-55$~keV) in the INCA FOV (top) and projected onto the equator along the line of sight (bottom). The dashed circles of the ENA equatorial projection are separated by 5~R$_S$.}
\label{figsup}
\end{figure*}

The third burst, which rises abruptly at about DOY 28 01$:$00~UT and lasts for about 5 hours, is distinguished from his neighbors by an enhanced peak emission and a spectrum which continuously extends toward unusually low frequencies, down to a few kHz (which corresponds to radio sources extending up to several planetary radii along auroral field lines). The particularly bright ENA event visible in the post-dawn sector at around 02$:$30~UT on DOY 28 in Figure \ref{fig1}d (arrow) or animation S$_2$, rising from the midnight sector a few hours before the SKR intensification, indicates a powerful injection of energetic ions. The temporal and LT correspondence between this ENA event and the UV and IR aurorae is highlighted in Figure \ref{figsup} with approximately simultaneous observations around 03$:$10~UT. As for the SKR recorded sources, they could not be located through goniopolarimetry because Cassini was too far from the planet. Instead, the radio sources visible from the spacecraft were determined using a simulation code detailed in section \ref{skr_modeling}. Figure \ref{figsup}d displays the locus of the SKR sources visible at that time (top) and their auroral footprint (bottom). In summary, the ENA bright region reached their maximal size and intensity when crossing the dawn meridian, coinciding with the peak of third SKR burst, on field lines connected to the active auroral region.

Similar radio signatures (enhanced intensity, low frequency extension) have previously been related to susbtorm-like activity driven by plasmoid release in the magnetotail \citep{Jackman_JGR_09}. Plasmoid ejections have also been related to energetic particle injections by \citet{Hill_JGR_08}, who reported the pass of a tailward propagating plasmoid associated with an ENA intensification in the inner magnetosphere. Therefore, the massive injection and the subsequent dawnside auroral activity associated with the third SKR burst may result from a significant plasmoid ejection \citep{Brandt_GRL_08,Carbary_JGR_08,Mitchell_PSS_09}. Interestingly, considering the colocation of SKR sources with the main atmospheric oval, the observed low frequency extension implies radio emission extending to larger distances from the planet along usual auroral field lines, and not to radio emission on field lines of higher latitudes, generally activated by solar wind compressions.

The other SKR bursts were less intense, but all well defined and phased with enhanced ENA emissions. This similar behavior observed for the whole interval suggests regular nightside injections at the S SKR phase for (at least) four consecutive rotations, with a variable intensity.


We checked solar wind parameters over the interval ranging from DOY 14 to 46, given by MHD heliospheric propagation from the Community Coordinated Modeling Center (not shown), with a typical uncertainty of a few days. Two significant solar wind compressions, with abrupt increases of dynamic pressure (from 0.01 to 0.12~nPa), were predicted around DOY 14 and 40. We identified powerful and long-lasting possible SKR counterparts within 1-2 days of each. Simulations then indicated quiet conditions from DOY 19 to 28, followed by a slow and modest third increase of dynamic pressure (from 0.01 to 0.06~nPa) peaking on DOY 32. We note that several episodic SKR enhancements similar to that of DOY 28 (for instance on DOY 20, 22, 26) occurred during the episode of quiet solar wind, while another powerful SKR burst on DOY 31 just preceded the peak of dynamic pressure evoked above. These considerations suggest that a solar wind compression was not responsible for the observed intensification of ENA and auroral activity. Instead, the latter may result from an internal trigger, such as a rotating partial ring current.


\subsection{Isolated features in sub-corotation}

On top of its general modulation, the SKR spectrum exhibits a variety of arc-shaped sub-structures in the time-frequency plane of Figures \ref{fig1}a and \ref{fig2}a. These features appear and disappear over timescales from hours to minutes, with either vertex-early (centre of curvature on the right) or vertex-late (centre of curvature on the left) shapes. \citet{Lamy_JGR_08b} previously showed that, for a fixed observer, a rotating field line populated with anisotropic CMI sources results in the observation of a complete arc (consisting of both its vertex-early and vertex-late edges, able to join at high frequencies), provided that the emission lasts for the whole time interval. In other words, the arc reveals the SKR activity of the small, time-varying, visible portion of the moving field line. The shape and temporal extent of the arc mainly depend on the beaming angle of radio sources and the rotation velocity of the field line. The authors modeled several SKR arcs, and identified the first example of a north/south conjugate sub-corotating SKR source. 

Assuming now that the large scale magnetic conjugacy observed between SKR sources and UV aurorae \citep{Lamy_JGR_09} also applies to localized sub-structures, the numerous radio arcs observed in Figure \ref{fig1}a can be expected to represent the signature of isolated active field lines connected to atmospheric auroral spots. In fact, most of these arcs are incomplete and restrict to part of either their vertex-late or vertex-early edge only, with typical durations ranging from $\le$1~h (DOY 27.2-27.4, early and late DOY 28) to several hours (double arc around DOY 27.5). Such durations mainly reflect the time-variable intensity of the source region. Radio observations thus provide a rich and direct diagnosis of isolated auroral features and their dynamics down to very short timescales.


The clearest SKR arc is visible at the beginning of DOY 28. Simultaneous UV (images d$_1$ to d$_5$) and IR (images 28 to 32) data are displayed in frames 28-32 of animation S$_1$, where LT$_{sc}$ is marked by green solid lines on polar projections. Considering a fixed spacecraft and sources moving in the direction of the planetary rotation, the vertex-early shape of the SKR arc requires an auroral source initially located west of the meridian plane of Cassini, moving toward the spacecraft. As a matter of fact, UV images of H$_2$ confirm the presence of a dawnside isolated spot rotating within the main oval, labelled Sp in Figure \ref{fig1}c, approximately visible from 06$:$00 (image d$_1$) to 07$:$30~LT (image d$_5$). This corresponds to 04$:$00 to 02$:$30~LT before LT$_{sc}$, a typical range for visible SKR sources \citep{Cecconi_JGR_09,Lamy_JGR_09}, which is investigated in more detail in section \ref{skr_modeling}. The intensity of the spot decreases with time by 25\% from $\sim$40~kR to $\sim$30~kR. A counterpart spot can also tentatively be seen in H and H$_3^+$ emissions in spite of lower SNR. Figure \ref{fig2}b shows that the spot Sp moves linearly in the (t,LT) plane along a dotted-dashed line throughout its lifetime. However, the slope of the line is smaller than that of rigid corotation, and indicates that the hot spot Sp sub-corotates at 65\% of the S SKR period. This matches the mean speed of magnetospheric cold plasma ranging equatorial distances from 10 to 15~R$_S$ \citep{Thomsen_JGR_10}. This findings reconciles the apparent contradiction between auroral corotational and sub-corotational dynamics, as both seemingly co-exist on adjacent magnetic field lines connected to the auroral oval.

We only had one clear example of an SKR arc associated with a UV spot over the interval where UV and IR data were available. However, this result completes the observation of numerous sub-corotating auroral spots on the one hand, numerous SKR arcs (some of which have been shown to sub-corotate) on the other hand, while the main auroral oval is known to be spatially conjugate with the bulk of SKR emission. These converging hints support a general conjugacy between SKR arcs and auroral spots.



\subsection{Secondary emissions}

Equatorward emissions appear to be nearly fixed in LT with a faint steady-state and homogenous intensity. However, as their latitude remains lower than that of the main oval, their position may be subject to rotational modulation as well. 

By contrast, high latitude emissions are highly intermittent and they generally vary from one image to the next, revealing temporal variability down to 15~min, without any obvious variation of their intensity nor locus with rotation. This matches a generation by solar-wind controlled acceleration processes on open field lines \citep{Badman_Icarus_11}.

\section{Energy budget}
\label{energy}

This section takes advantage of our set of remote observations to provide information on the magnetosphere-ionosphere coupling in terms of energy transport and dissipation. 

\subsection{Atmospheric precipitation}

\subsubsection{Primary electrons}
\label{prim_elec}

\begin{figure*}
\centering\includegraphics[width=1.\textwidth]{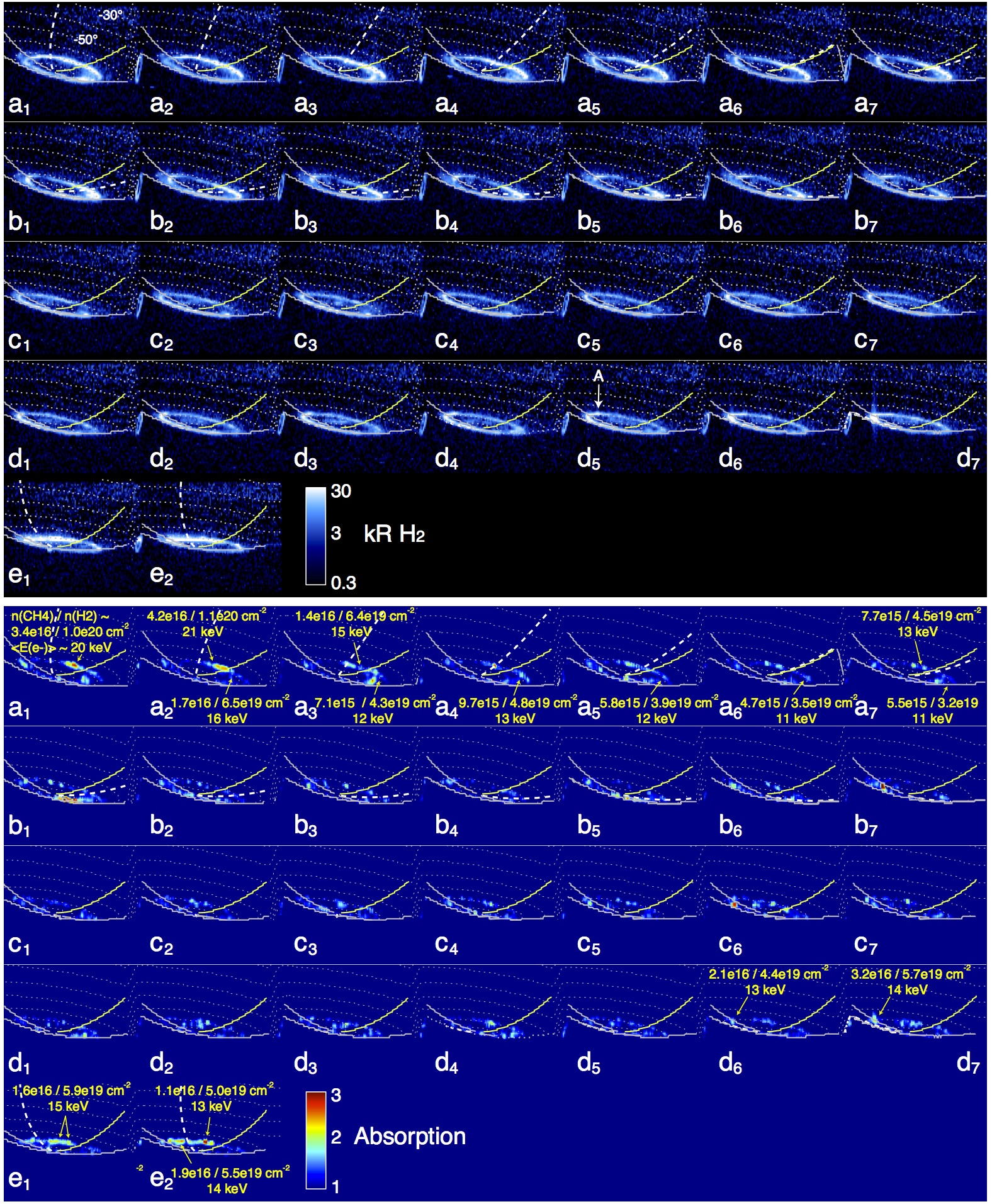}
\caption{(Top) Brightness of total H$_2$ corrected for absorption. (Bottom) Absorption factor smoothed over 3 pixels once pixels with low SNR (large background values, weak H$_2$ emission, pixels beyond the limb) have been removed. Absorption factors $\ge1$ indicate absorption. Yellow numbers indicate the slant (observed) CH$_4$ column density above the auroral source along the line-of-sight, the vertical H$_2$ column density (derived from the equatorial atmospheric model of \citet{Moses_Icarus_00} adapted to the latitude of 75$^\circ$) and the primary energy of electrons derived from direct fitting of the absorbed spectra \citep{Gustin_Icarus_09}, whose uncertainty is estimated to 20\%.}
\label{fig3}
\end{figure*}

FUV spectroscopy is a powerful tool to diagnose auroral precipitations. The neutral upper atmosphere of Saturn is populated with hydrocarbons, mostly CH$_4$, whose absorption cross-section is roughly constant from 110 to 130~nm and rapidly drops at larger wavelengths \citep{Au_CP_93}. When FUV auroral photons travel through a CH$_4$ layer on their way out, they undergo absorption below 135~nm, following the Beer-Lambert law. One simple way to quantify it is to derive a color ratio (CR), defined by the ratio between the intensities integrated over spectral ranges including unabsorbed and absorbed bands, respectively \citep{Yung_ApJ_82,Livengood_these_91,Rego_JGR_99}. Once normalized by the color ratio calculated for a reference unabsorbed spectrum, CR~$\ge1$ indicate absorption. In appendix \ref{uvis_processing}, we directly (and equivalently) derived the absorption by the ratio of H$_2$ model spectra fitted to 123-130~nm and 144-166~nm bands. Results are plotted for all UVIS images in Figure \ref{fig3} (bottom). We find non negligible absorption all along the auroral oval with typical values from 1 to 3, peaking within the active rotating region (images of sequences a, d or e), for some nightside emissions at the limb (images b$_1$, b$_5$), and possibly within the polar region (images d$_2$, d$_3$). Conversely, we note that in some bright regions of the main oval, the emissions are not absorbed (images of sequences b and c). This indicates a non linear relationship between the radiated flux and the absorption, which is investigated in more detail below.

The absorption factor can be used to determine the CH$_4$ column density above the emission peak along the slant path and its zenithal value once corrected by the viewing geometry. This can then be translated into an H$_2$ zenithal column density using a model atmosphere \citep{Moses_Icarus_00}, which eventually gives the penetration depth of the precipitating electrons, and, in turn, their primary energy (see for instance \citep{Galand_JGR_11}). Absorption remains controlled by the altitude of CH$_4$ layers and, in the case of Saturn, it is generally detectable only for electron energies larger than 10~keV. To accurately determine the energy of the precipitating electrons, we use the technique described by \citet{Gustin_Icarus_09}, based on this principle, but directly fitting observed spectra over 126-165~nm. The oblique (observed) CH$_4$ column densities, the zenithal H$_2$ column densities, and the energies of primary electrons labelled on Figure \ref{fig3} (bottom) were derived from average spectra built for a set of localized bright regions. The slant CH$_4$ and zenithal H$_2$ column densities vary within $5\times10^{15}-5\times10^{16}$~cm$^{-2}$ and $3\times10^{19}-1\times10^{20}$~cm$^{-2}$ respectively.

This leads to precipitating electron energies typically within the range $10\pm2$ to $20\pm4$~keV, and modestly LT dependent : $15-20$~keV around noon (a$_1$-a$_3$), $10-15$~keV at dusk (a$_3$-a$_7$) and $\sim$$15$~keV from dawn to noon for the auroral enhancement ($d_6$-$e_2$). However, the absorbed regions well match the location of the active rotating region. Polar emissions are not significantly absorbed, an indication of lower electron energies, likely equal or less than 10~keV. Finally, absorbed emissions at the limb could not provide any quantitative estimate, as the emission angles are poorly known, resulting in unreliable zenithal column densities. Nonetheless, they indicate bright emission on the nightside, able to cross a large portion of the atmosphere along the limb.


\subsubsection{Power input in the atmosphere}
\label{power_input}

\begin{figure}
\centering\includegraphics[width=0.48\textwidth]{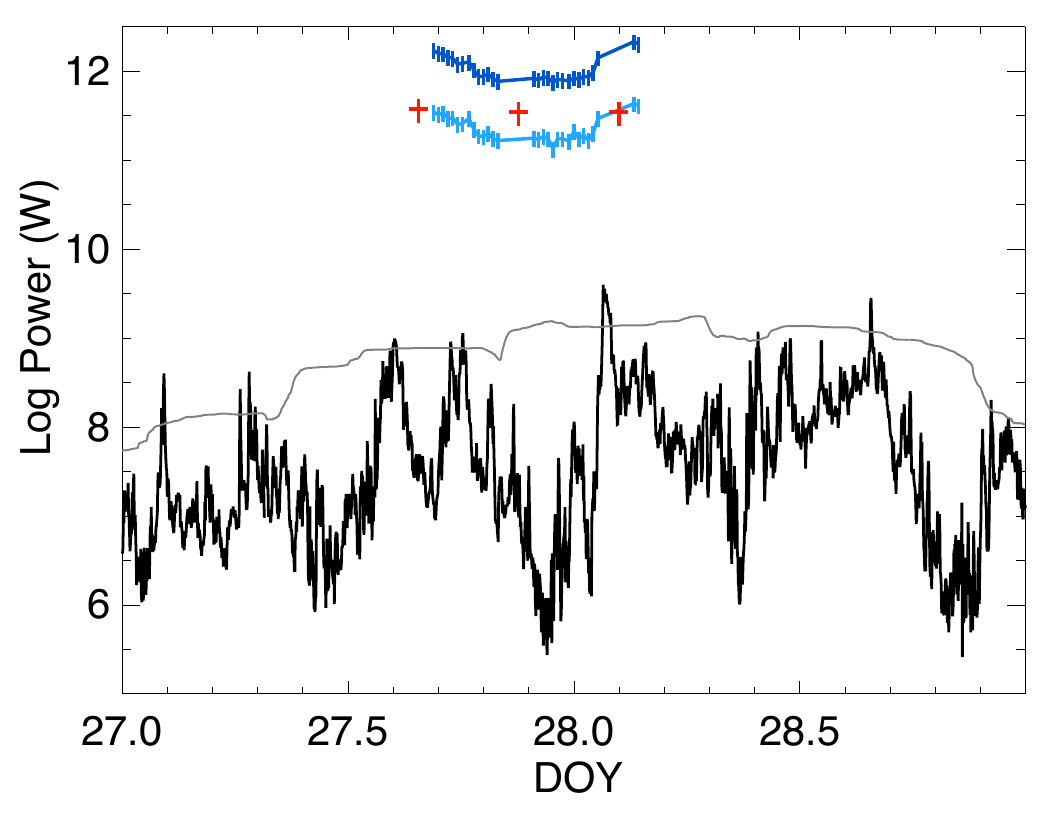}
\caption{Total precipitated power as a function of time (dark blue, as derived in section \ref{power_input}) and auroral power radiated by H$_2$+H Ly-$\alpha$ (light blue, see section \ref{power_H_H2}), H$_3^+$ (red, see section \ref{power_H3p}) and SKR (black and gray, see section \ref{power_SKR}).}
\label{fig4}
\end{figure}

The total input power carried by precipitating electrons can be estimated from the total brightness of H$_2$ corrected for absorption (as radiated by the source, and hereafter simply referred to as the brightness of H$_2$) and the conversion factor of $\sim$10~kR radiated per incident mW.m$^{-2}$ \citep{Gerard_JGR_82,Waite_JGR_83}. The input powers are plotted as dark blue crosses in Figure \ref{fig4}. They vary between $7.4\times10^{11}$ and $2.1\times10^{12}$~W ($8.6\times10^{11}$~W median), with an enhancement by a factor of 3 from image d$7$ to e$_1$. These values match and even slightly exceed the upper range of the $1-10\times10^{11}$~W previously estimated for S aurorae observed in January 2004, which exhibited a powerful brightening induced by a solar wind compression \citep{Clarke_Nature_05}. The uncertainty on the input power is estimated to 20\% (appendix \ref{uvis_processing}).


The bright rotating region in Figure \ref{fig3}a can reach brightnesses as high as $170$~kR (images e$_1$-e$_2$), well above the $5-10$~kR of the rest of the main oval. This corresponds to local energy inputs varying from $\le1$ to 17~mW.m$^{-2}$. 

Looking at the relationship between the brightness of H$_2$ (Figure \ref{fig3}a) and the energy of primary electrons (Figure \ref{fig3}b) for the bright rotating region, we note two remarkable trends. For the peculiar enhancement visible from image d$_6$ to e$_2$, the electron energy is only modestly varying (by a factor of $\sim$1.2) while the associated brightness varies much more (by a factor of $\ge2$). This suggests that large auroral intensifications are less controlled by the increase of electron energy than by the increase of their flux. Conversely, in the more usual situation seen from image a$_1$ to a$_7$, the electron energy and the brightness of the bright region vary in a very similar way. This indicates that, in this case, the electron energy appears as the main trigger of brightness variations.

\subsubsection{Current system}

Several FAC systems have been proposed to account for the observed auroral precipitations and their rotational modulation \citep[and refs therein]{Andrews_JGR_12}. As an attempt to interpret the close relationship between bright ENA and auroral sources, further confirmed here, \citet{Mitchell_PSS_09,Brandt_GRL_10} proposed FAC driven by plasma pressure gradients within a rotating partial ring current, with ENA sampling the singly charged energetic ions, and therefore the hot plasma pressure contribution. The uncertainty of ENA flux was nonetheless too large to provide quantitative estimates of the FAC strength. 

The spatial distribution of the ENA bright event, however, indicates that the FAC peaks over a broad longitudinal sector coincident with the auroral active region. This does not fit the picture of large-scale FACs generated on each side of the longitudinally asymmetric proton distribution, where the pressure gradients are maximum. The observed ENA morphology rather supports the co-existence of multiple filamentary FAC structures associated with small-scale regions of hot plasma \citep{Mitchell_PSS_09}, which may by initiated by pressure gradients or another trigger. INCA observations with better spatial resolution remain necessary to address this question.

\subsection{Auroral radiated power}

\subsubsection{Neutral atmosphere}

\subsubsubsection{Relative contributions of H Ly-$\alpha$ and H$_2$ bands}

\begin{figure*}
\centering\includegraphics[width=1.\textwidth]{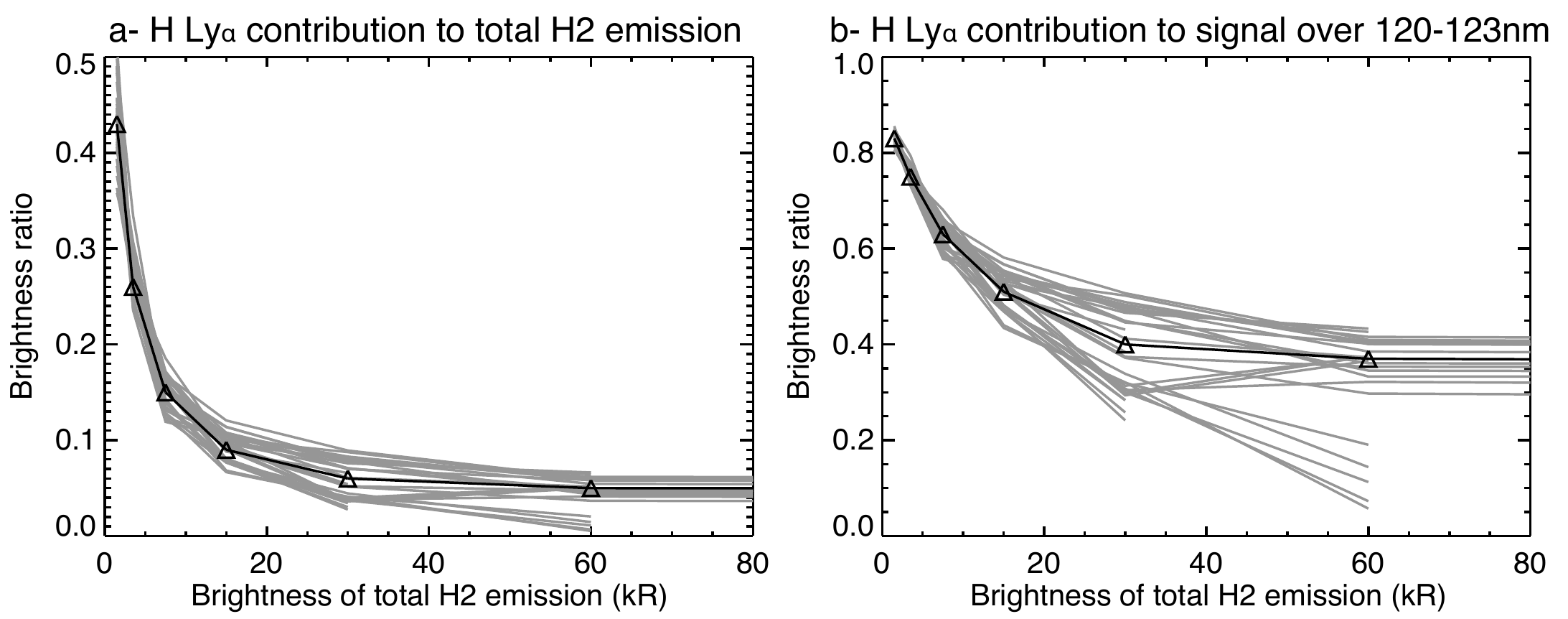}
\caption{Brightness ratios of H Ly-$\alpha$ to (a) total H$_2$ emission over 80-170~nm and (b) total H and H$_2$ emission over 120-123~nm, all being corrected for hydrocarbon absorption, as a function of the brightness of H$_2$. The ratios were calculated for average spectra built from all the auroral pixels of UVIS images a$_1$ to e$_2$ taken separately (gray curves) or taken together (black curves and triangles) for increasing ranges of H$_2$ brightness. The values corresponding to the black triangles are listed in table \ref{tab1}.}
\label{fig_ratios}
\end{figure*}

Figure \ref{fig_ratios} quantitatively displays the contribution of the H Ly-$\alpha$ line to the total brightness of H$_2$ bands (panel a) and to the total brightness measured over 120-123~nm (panel b) as a function of the brightness of H$_2$, corrected for atmospheric absorption. Both brightness ratios were calculated for average spectra built from the auroral pixels of all UVIS images a$_1$ to e$_2$ taken separately (gray curves) or merged together (black curves and triangles) for increasing ranges of H$_2$ brightness. Table \ref{tab1} lists the brightness ratios and associated absorption factors (as defined in section \ref{prim_elec}) corresponding to the black triangles.


Overall, both brightness ratios strikingly decrease with increasing brightnesses of H$_2$ similarly for all images (gray curves). The small dispersion of the individual plots strengthens the reliability of the average plot (black curves), and hence an increased Ly-$\alpha$ contribution for faint emissions. The black triangles in Figure \ref{fig_ratios}a (\ref{fig_ratios}b) lie below $10\%$ ($50\%$) for H$_2$ brightnesses above 10~kR and reach $45\%$ ($82\%$) for 1-2~kR, leading to an average $13\%$ ($58\%$) ratio. In addition, Figure \ref{fig_ratios}b highlights the contamination of the H Ly-$\alpha$ line by H$_2$ emission, and justifies the need to correct for it. 

As H Ly-$\alpha$ and H$_2$ emission excitation efficiencies vary with altitude, their relative contribution directly relates to the penetration depth, and in turn the energy of primary electrons. The measurement of brightness ratios thus provide a test of electron energies complementary to FUV spectroscopy. Concerning the  H Ly-$\alpha$/H$_2$ ratio (Figure \ref{fig_ratios}a), our average estimate of 13\% compares to the theoretical values of 13\% \citep{Shingal_JGR_92} or 11\% \citep{Rego_JGR_94} calculated under unabsorbed conditions for 10~keV electrons. This energy thus appears as a reliable estimate for typical electrons responsible for the main oval. Since 10~keV also corresponds to the lower limit for which FUV absorption is detectable, this explains why Saturn's FUV spectra generally display little absorption \citep{Gustin_Icarus_09}. 

\citet{Rego_JGR_94} predicted that lower energy electrons are expected to increase the H Ly-$\alpha$ relative contribution to the spectrum. A preliminary study at the H Ly-$\alpha$/H$_2$ brightness ratio in images with significant polar emissions (sequences c and d) indicate mean ratios which significantly exceed the 13\% value discussed above. This suggests that electrons precipitating in the polar cap must be less energetic ($\le10$~keV). 

This approach may bring more quantitative estimates, particularly below 10~kR, where brightness ratios are extremely sensitive. Indeed, the trend observed for the black lines in Figure \ref{fig_ratios} (increasing ratios toward lower brightnesses) supports the use of the brightness of H$_2$ as a rough proxy of the electron energy, as independently shown above. In the previous section, we noticed a poor correlation between the electron energy and the brightness of H$_2$ as derived for a set of spatially localized regions. However, we here remark that the absorption, as given in table \ref{tab1} for average spectra, clearly increases with the brightness of H$_2$ beyond 5~kR on the one hand. This is all the more remarkable that average spectra merge spatial pixels distributed over the whole auroral region, and thus very different incidence angles. On the other hand, a change of the electron flux itself shall not affect the brightness ratio. This method could be checked by investigating any relationship between high brightness ratios (low brightnesses of H$_2$) and altitudes of emission measured on images (as done for intense aurorae \citep{Gerard_GRL_09}), but this is beyond the scope of this paper.

\begin{table*}
\center
\begin{tabular}{c|c|c|c|cl}
  Brightness of H$_2$&Individual spectra&Absorption&H Ly-$\alpha$/H$_2$&H Ly-$\alpha$/band 120-123~nm\\
  \hline
  1-2~kR&6452&$1.18\pm0.12$ &43\%&83\% \\
  \hline
  2-5~kR&6956&$1.13\pm0.12$&26\%&75\% \\
  \hline
  5-10~kR&3751&$1.12\pm0.12$&15\%&63\%\\
  \hline
  10-20~kR&2968&$1.20\pm0.12$&9\%&51\%\\
  \hline
  20-40~kR&1545&$1.24\pm0.13$&6\%&40\%\\
  \hline
  40-80~kR&449&$1.32\pm0.14$&5\%&37\%\\
  \hline
  $\ge80$~kR&159&$1.81\pm0.19$&4\%&30\%\\
  \hline
  \hline
  $\ge$1~kR&22335&$1.26\pm0.13$&12\%&58\%\\
\end{tabular}
\caption{Characteristics of average spectra built from all auroral pixels taken together for the ranges of H$_2$ brightness given in the left column : number of individual spectra, absorption factor (as defined in section \ref{appendix_absorption}), and ratios of H Ly-$\alpha$ to H$_2$ emission over 80-170~nm and to total H + H$_2$ emission over 120-123~nm corresponding to the black lines of Figure \ref{fig_ratios}.}
\label{tab1}
\end{table*}

\subsubsubsection{Radiated power}
\label{power_H_H2}

The total power emitted by auroral H$_2$ and H Ly-$\alpha$ once corrected for absorption is obtained as described in appendix \ref{uvis_processing}. The resulting values are plotted in light blue on Figure \ref{fig4}, and vary within the range $1.3-4.3\times10^{11}$~W ($2.0\times10^{11}$~W median), significantly larger than the 2-30~GW obtained in January-February 2007 and February 2008 \citep{Clarke_JGR_09}. The latter authors corrected for the viewing geometry of the southern pole, so that the discrepancy of average powers likely comes from different calculations : previous estimates may be underestimated by a restriction to the HST FUV bandpass \citep{Gustin_JMS_13} and/or the use of a model auroral spectrum, generally unabsorbed, while we here derived the real total radiated power separately for each pixel. Compared to the input power, this yields a median 21\% efficiency (3\% for H Ly-$\alpha$ emission alone).

The main oval is the predominant source of radiated power, with equatorward emissions 1-2~kR bright, considered hereafter as negligible. To estimate the contribution of polar emissions, we derived the ratio of intensities integrated from $-73^\circ$ to $-85^\circ$ (polar) and from $-68^\circ$ to $-73^\circ$ (main oval) between 06$:$00 and 13$:$00 LT for images of sequences c and d, where significant polar aurorae were observed (frames 21-34 in animation S$_1$). This sequence has the advantage of displaying a roughly constant auroral morphology, easily discriminated by the above ranges, and observed simultaneously in UV and IR. The results are plotted in Figure \ref{fig5}. Despite a noisier H Ly-$\alpha$ emission (see appendix \ref{uvis_processing}), the polar-to-main-oval emission ratio is significant for both H Ly-$\alpha$ and H$_2$, with median values of $0.43\pm0.27$ and $0.19\pm0.07$ respectively. These are mainly indicative, as both types of emission originate from different processes and our analysis restricts to a particular interval, when polar emissions were bright. Nonetheless, they still give an order of magnitude when both emissions are observed together, and, more importantly, can be compared between species. In particular, the larger ratio obtained for  H Ly-$\alpha$ is consistent with the results of the above section, and suggests that the electrons precipitating in the polar region are less energetic than those precipitating along the main oval.

\begin{figure}
\centering\includegraphics[width=0.48\textwidth]{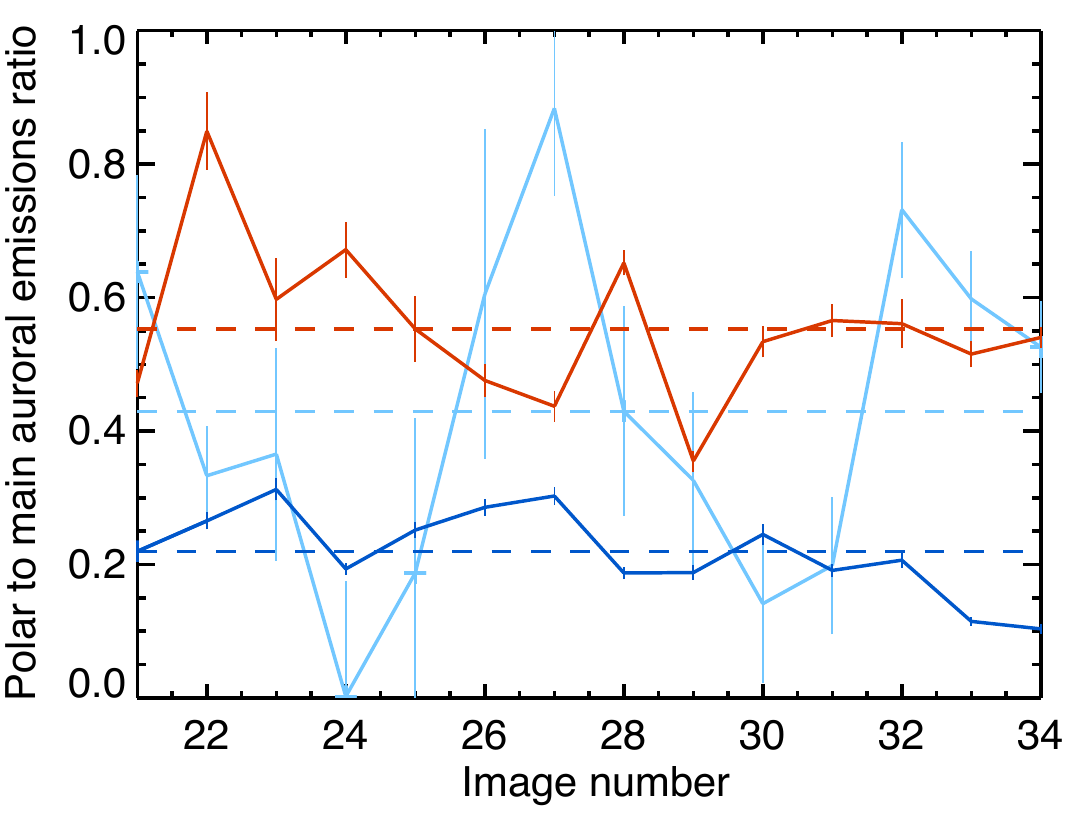}
\caption{Ratio of auroral emissions integrated over $-68^\circ$ to $-73^\circ$ and $-73^\circ$ to $-85^\circ$ between 06$:$00 and 13$:$00 LT for H-Ly-$\alpha$ (light blue) and H$_2$ (dark blue) images c$_1$ (numbered 21) to d$_7$ (numbered 34), and H$_3^+$ (red) images 21 to 34. Statistical errrors are indicated by vertical bars for each image. Median values of $0.43\pm0.25$, $0.22\pm0.07$ and $0.55\pm0.12$ and respectively are indicated by colored dashed lines.}
\label{fig5}
\end{figure}

\subsubsection{Ionosphere}

\subsubsubsection{Temperatures}
\label{temperatures}

The ionospheric H$_3^+$ emissions being temperature-dependent, any power estimate first requires the determination of the temperature. Due to the low SNR of the majority of VIMS images, we restricted our analysis to long-integration time observations (images 3, 20 and 37). We assumed Local Thermodynamic Equilibrium (LTE) conditions as the kronian $H_3^+$ emission peak lies at altitudes where LTE is valid \citep{Tao_Icarus_11}. We then derived the temperature as described by \citet{Melin_GRL_11}, and discussed in more details in appendix \ref{vims_processing}. 

Figure \ref{fig6} shows the results of this analysis, with temperature spatial maps (second row) and plots as a function of latitude (third row) and LT (bottom row). Only pixels with sufficient SNR are plotted, so that solely the auroral region (including equatorward and polar emissions) is covered in practice. Interestingly, the temperatures do not vary significantly from image to image or as a function of latitude nor LT in spite of time-variable fluxes (top row). Only the pre-dawn portion of the main oval of image 20 seems to display a localized temperature increase exceeding the 3$\sigma$ level above the median temperature, but a further inspection of the corresponding H$_3^+$ intensities show faint emission, so that we cannot consider this variation as reliable. Overall, and in spite of a large dispersion, the temperature is not clearly related to bright emitting regions (see also Figure \ref{fig_app2}) or hard precipitations identified above along the main oval. Instead, the temperature rather appears uniform across the auroral region, suggesting an efficient distribution of auroral heating. As a result, H$_3^+$ molecules are permanently thermalized and the emission is controlled by auroral precipitations. Median values for the 3 images are $420\pm85$,  $430\pm88$ and $392\pm98$~K respectively, consistent with the $440\pm50$~K previously derived for one observation of the main oval \citep{Melin_GRL_11}.

\begin{figure}
\centering\includegraphics[width=0.5\textwidth]{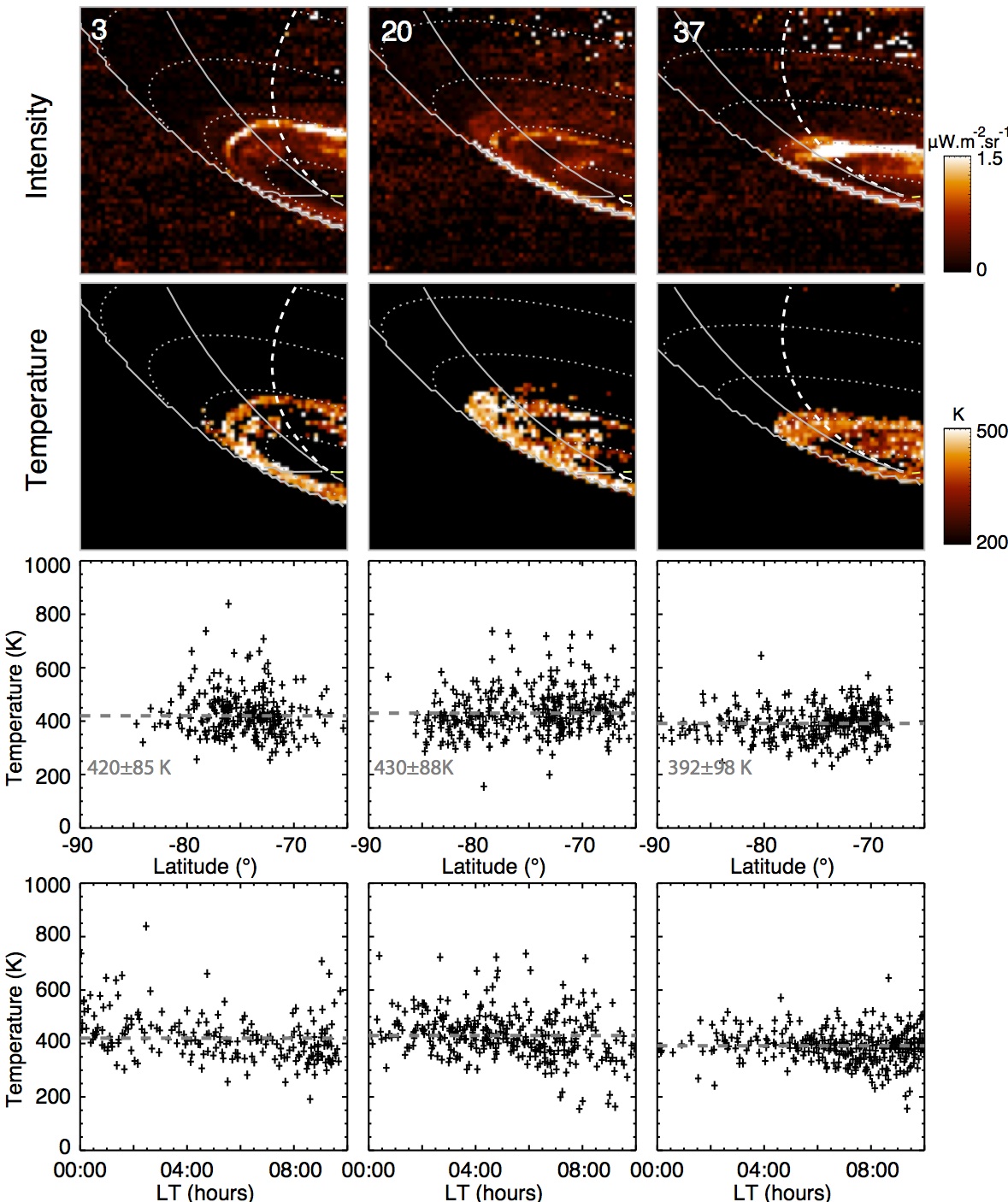}
\caption{(First row) Maps of H$_3^+$ flux (images 3, 20 and 37) in a format similar to animation S$_1$. (Second row) Corresponding temperature maps, with low SNR pixels removed. (Third row) Ionospheric temperature per pixel as a function of latitude. (Last row) Ionospheric temperature per pixel as function of LT. Gray dashed lines indicate median values.}
\label{fig6}
\end{figure}

\subsubsubsection{Radiated power}
\label{power_H3p}

The H$_3^+$ aurorae differ only slightly from the UV ones. As stated above, they exhibit coincident morphological features. But contrary to the H-Ly-$\alpha$ and H$_2$ cases, the intensity of the H$_3^+$ main oval compares to that of polar emissions, the equatorward emissions being, again, negligible. The polar-to-main-oval ratio has been estimated for VIMS images 21-34 similarly to UVIS ones. The result is plotted by the red curve in Figure \ref{fig5}, with values varying {beyond error bars} between 0.5 and 0.75, and decreasing with time in the same way as the ratio of H$_2$. This leads us to consider that the median H$_3^+$ ratio of $0.55\pm0.12$ is meaningful, and that the polar H$_3^+$ emission is slightly brighter relative to the main oval than for the neutral species. We note that this value is qualitatively close to that previously derived by \citet{Badman_Icarus_11} (0.84) even if the latitudinal ranges used to separate polar and main regions were different.

The H$_3^+$ column density and total power radiated by the whole S auroral region were then estimated for the three long-integrated VIMS images, as described and discussed in appendix \ref{vims_processing}. The column density per pixel reaches $10^{16}$~m$^{-2}$ in average (a few $10^{17}$~m$^{-2}$ peak), which compares to the previous estimates of \citet{Melin_GRL_11}. In terms of total radiated power, we found 3.7, 3.3 and $3.5\times10^{11}$~W for images 3, 20 and 38 respectively, with an uncertainty estimated to $\sim$~30\%. These values are precisely comparable to those estimated in section \ref{power_H_H2} and establish that the aurorae studied here radiate comparable power in the IR H$_3^+$ lines and in the UV H-Ly-$\alpha$ and H$_2$ bands. Although the morphology of the IR images clearly differ (Figure \ref{fig1}b, animation S$_1$), the comparable power values result from (slight) changes in temperature and (larger) changes of geometric corrections accounting for the portion of the auroral region not visible by VIMS. 

These values are compared to the precipitated powers in Figure \ref{fig4}. Quantative estimates can tentatively be done although the VIMS long-integrated images were not obtained simultaneously to UVIS measurements, so that the time variation of the aurorae from one long-integrated VIMS image (60~min) to the next UVIS image is neglected. We found emission efficiencies of 22, 42 and 16\% respectively.

\subsubsection{Radio power}
\label{power_SKR}

The SKR power was derived from RPWS calibrated measurements, once integrated between 3 and 1000~kHz and neglecting the contribution of narrowband emissions observed around 20~kHz during DOY 27. It is expressed in W.sr$^{-1}$ (black line in Figure \ref{fig4}), as the solid angle filled in by the radiation is unknown {\it a priori} \citep{Lamy_JGR_08a}. Retrieved values can reach a few 10$^9$~W.sr$^{-1}$ with a median value (50\% occurrence level over the whole interval) of $3.1\times10^7$~W.sr$^{-1}$. This compares to statistical estimates obtained from 2004 to late 2006, and indicates quiet conditions.

The SKR emission is strongly anisotropic, so that only a limited part of all existing radio sources can be observed at the same time. In a previous study, \citet{Kurth_Nature_05} investigated the temporal conjugacy of radio and UV auroral powers in Jan. 2004. They chose to rotationally-average the SKR power and derive a linear correlation coefficient of 0.87. We note that the (sub-)corotational motion of auroral features evidenced in this study justifies this approach {\it a posteriori}, as all radio sources lasting long enough are expected to illuminate the RPWS antenna during a rotation, so that the visibility factors are averaged. However, this is not applicable on timescales shorter than a planetary rotation.

For this reason, we first look for any correlation between the instantaneous SKR and total UV power radiated throughout the latter dataset. We calculated a rank correlation coefficient, more general than a linear one. Indeed, no linear relationship is expected between the intensity of radio and UV radiations, controlled by different physical parameters, namely the shape of the electron distribution function in the phase plane and the precipitating electrons, respectively. We eventually find a good coefficient of 0.74 (with a false alarm probability of $3\times10^{-6}$). 

In an effort to improve the conditions of the comparison, we restrict the UV power to that radiated by the main oval within the LT range where SKR sources are visible (ranging from 02$:$30 to 04$:$00~LT on both sides of the sub-spacecraft meridian, as discussed in the next section). The correlation coefficient increases up to 0.77. If we furthermore average the SKR power over the time needed for the corotating radio sources to sweep one of the 1.5~LT wide visible sectors, namely 40~min, so that the whole spectrum of radio sources populating a single field line crossing this region can be observed, this factor finally reaches 0.78 (with a false alarm probability of 3$\times10^{-7}$). This correlation is highly significant and strengthens the close relationship between SKR and UV aurorae.

In order to estimate the total SKR radiated power, we need to account for the anisotropy of the emission. Anticipating the results of section \ref{skr_modeling}, we use a typical instantaneous beaming angle of $60\pm10^\circ$, a few degrees thick, for all frequencies. Integrated over a rotation, the covered solid angle thus reaches $(1\pm0.3)\times2\pi$~sr. Once applied to the rotationally-averaged instantaneous SKR power, we estimate the total SKR radiated power, plotted in gray on Figure \ref{fig4}. It varies from 10$^8$ to $2\times10^9$~W, with a median value (50\% occurrence level) of $7.8\times10^8$~W, $i.e.$ $\sim$~$0.1\%$ of the median input power. 

This ratio shall be considered carefully. First, it relies on the assumption that the lifetime of radio sources is sufficiently long so that they can be detected over a rotation. This is not true for most of isolated short-lived arcs. Second, the input power refers to precipitating particles, while radio emission is triggered by unstable electron distributions, including non precipitating particles. For these reasons, $0.1\%$ rather gives a lower limit of the intrinsic electron-to-wave conversion efficiency, likely closer to the $1\%$ level previously calculated in situ \citep{Lamy_JGR_11}.

\subsubsection{Contribution of visible and X-ray aurorae}

The above energy budget is not complete as it does not deal with possible contributions of visible and X-ray aurorae, nor the energy loss through atmospheric heating.

Kronian visible aurorae have been clearly detected with Cassini \citep{Dyudina_AGU_07}. Visible observations were acquired during DOY 27-28 by the Imaging Science Subsystem (ISS), but the contrast with the reflected solar light was too large to investigate any auroral signal (U. Dyudina, personal communication). In terms of intensity, a brightness of about 10~kR was derived for a single example over the 300 to 900~nm range \citep{Kurth_09}. Alternately, \citet{Cook_JGR_81} estimated for Jupiter that visible emissions between 400 and 600~nm (Balmer series) display brightnesses approximately one order of magnitude lower than that  of H$_2$. As visible photons are less energetic than UV ones, the contribution of visible aurorae to the total dissipated power is expected to be minor.

X-ray aurorae have not been detected yet at Saturn in spite of several attempts with Earth-based telescopes. The most recent upper limits were estimated to a few tens of 10$^6$~W for charge-exchange and bremsstrahlung processes \citep[and refs therein]{Branduardi_JGR_13}. Their contribution to the radiated power is thus negligible.

\section{Modeling of SKR visibility}
\label{skr_modeling}

In this section, we focus on the analysis of SKR visibility through simulations of visible sources and dynamic spectra. 

For this purpose, we used a simulation code of the PRES (Planetary Radio Emissions Simulator) family, previously employed for both Jupiter \citep{Hess_GRL_08} and Saturn \citep{Lamy_JGR_08b}. Briefly, the code calculates whether a beam of CMI-driven emission, radiated by a radio source at the local cyclotron frequency and propagating in straight line from the source to the observer, is visible from the Cassini spacecraft. We used a standard magnetic field model composed of the Saturn-Pioneer-Voyager (SPV) model \citep{Davis_JGR_90} completed by a simple ring current \citep{Connerney_JGR_83}. Active magnetic field lines are populated with individual radio sources at frequencies ranging the observed SKR spectrum. Each radio source radiates waves within a conical sheet of emission, defined by an aperture angle $\theta$(f)~=~({\bf k},{\bf B}), {\bf k} and {\bf B} being the wave vector and the magnetic field vector at the source, and a thickness $\Delta\theta$. In section \ref{anisotropy}, the location of the source is fixed and the free parameter is $\theta$(f), while in sections \ref{visible_sources},\ref{dynamic_spectra}, the the cone aperture is taken from section \ref{anisotropy} and the free parameter is the location of the source region.

\subsection{Anisotropy}
\label{anisotropy}

\begin{figure}
\centering\includegraphics[width=0.48\textwidth]{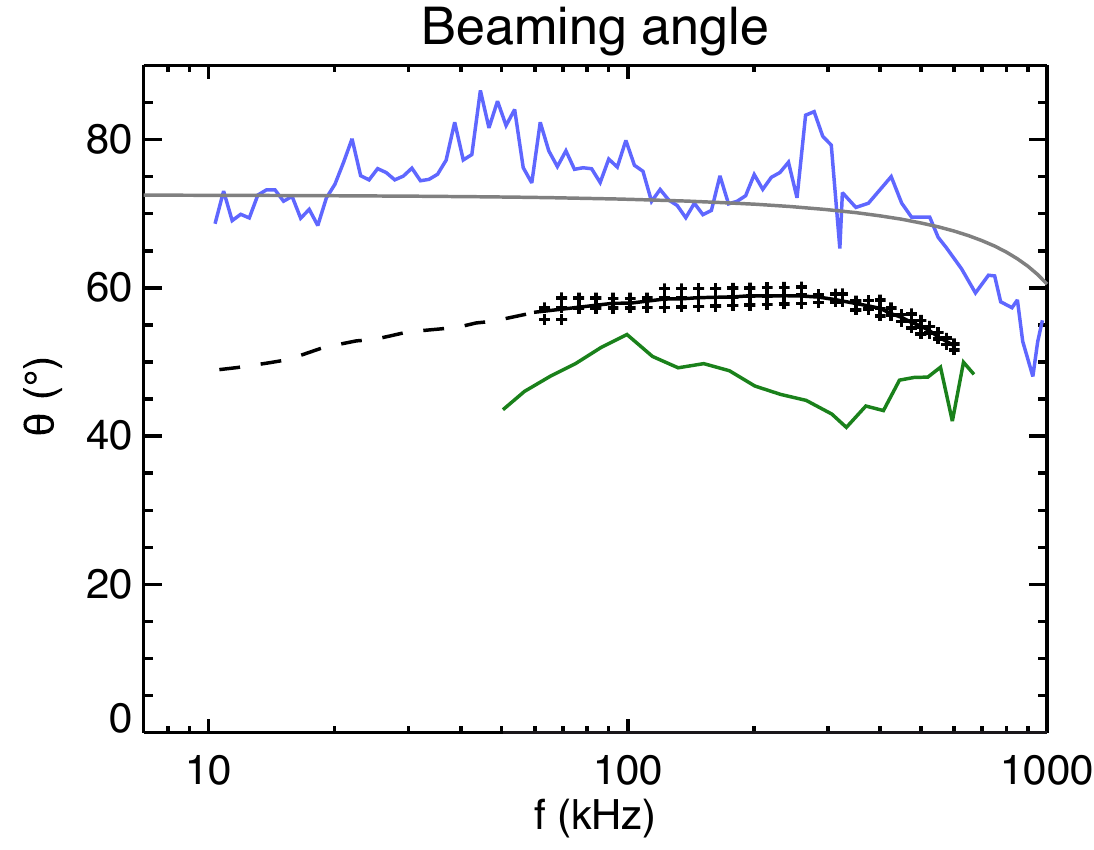}
\caption{SKR beaming angle $\theta(f)$ = ({\bf k},{\bf B}), where {\bf k} and {\bf B} are the wave vector (assuming straight line propagation) and the magnetic field vector at the source. The black crosses display the beaming angle derived for radio sources responsible for the SKR arc associated with the atmospheric hot spot from DOY 27.98 to 28.05 (both labelled Sp in Figure \ref{fig2}). Their average is plotted with the black solid line. The temporal width of the arc yields a dispersion of black crosses of a few degrees. The black dashed line plots a low frequency extrapolation. The other colored lines indicate, for comparison, the beaming angles derived from previous modeling (gray line \citep{Lamy_JGR_08b}) and observational (green line \citep{Cecconi_JGR_09} and blue line \citep{Lamy_JGR_11}) studies.}
\label{fig7}
\end{figure}

In section \ref{sources}, we identified a UV spot in sub-corotation within the main auroral oval associated with a well defined SKR arc (both labelled Sp in Figures \ref{fig1}c,\ref{fig2}a,b). Taking advantage of the colocation of UV and radio emissions along the same flux tube, supported by a previous statistical study \citep{Lamy_JGR_09}, we directly derive the SKR beaming in a self-consistent manner. We assume a single active magnetic field line connected to each successive location of the atmospheric hot spot and we determine $\theta_{obs}$(f) for each visible frequency of the arc. The individual determinations of the beaming angle are plotted as a function of frequency by black crosses in Figure \ref{fig7}. The temporal (horizontal) width of the arc yields a (vertical) dispersion of $\theta_{obs}$(f) by a few degrees, taken as a typical uncertainty on $\theta_{obs}$(f). The average beaming angle is shown by the black line in Figure \ref{fig7}, together with a low frequency extrapolation. Values derived from previous studies are superimposed with colored lines for comparison. 

The spectral evolution of $\theta_{obs}$(f) is roughly consistent with past determinations, with a common decrease at high frequencies. Nonetheless, beyond 100~kHz, where refraction effects are neglected and where the assumption of straight line propagation is most reliable, $\theta_{obs}$(f) lies between $50^\circ$ and $60^\circ$, inbetween the previous determinations by $\sim~10^\circ$. We previously proposed that this discrepancy, whose analysis is beyond the scope of this paper, could result from either temporal and/or azimuthal variations of the beaming angle. For our purpose, it is important to note that the derived $\theta_{obs}$(f) already accounts for any possible refraction close to the source.

Finally, the temporal extent of the observed vertex-early SKR arc can be translated into a LT range swept by the auroral spot S, which is indicated by the top pair of green lines in Figure \ref{fig2}. This range, extending from 02$:$30 to 04$:$00~LT before LT$_{sc}$, directly defines the region of visible radio sources west of the meridian plane of Cassini. Assuming constant SKR visibility conditions over the lifetime of the auroral spot, a second region of visible radio sources can be symmetrically defined east of the meridian plane of Cassini by 02$:$30 to 04$:$00~LT after LT$_{sc}$, as indicated by the bottom pair of green lines in Figure \ref{fig2}. 

\subsection{Visible sources}
\label{visible_sources}

Once the SKR beaming angle is determined, it is straightforward to model which radio sources are visible by Cassini. Hereafter, we assume radio sources throughout the observed SKR spectrum, ranging from 10 to 800~kHz, the beaming angle $\theta_{obs}$(f) determined above together with a $\Delta\theta=5^\circ$ thickness \citep{Lamy_JGR_08b} (identified as an upper limit in \citep{Lamy_JGR_08b}). 

\begin{figure*}
\centering\includegraphics[width=1.0\textwidth]{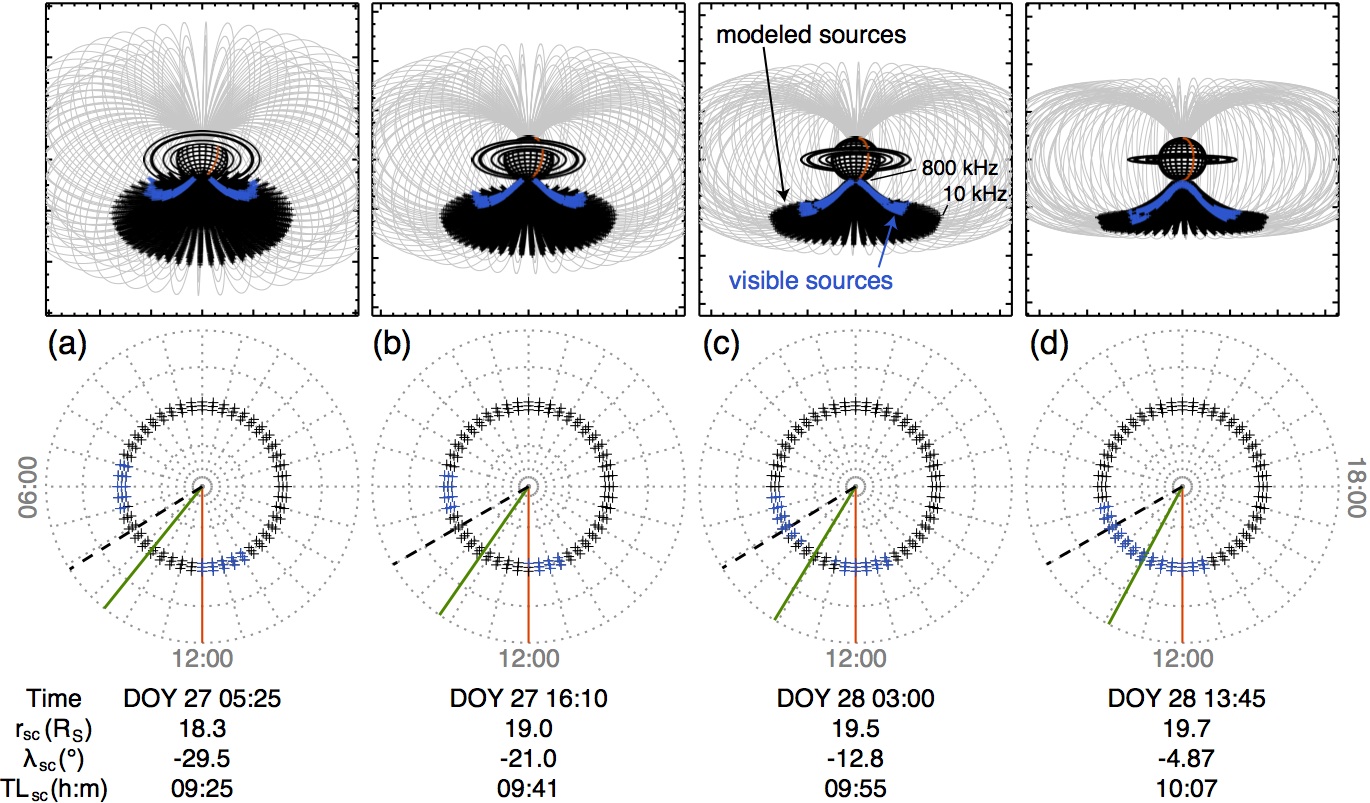}
\caption{Simulations of the radio visibility for the four SKR predicted peaks during DOY 27-28, as seen from the spacecraft (top) and as polar projections (bottom). The footprint of the source region is approximated by a circular oval extending from $-71^\circ$ to $-69^\circ$ latitude (crosses in the bottom panel). Magnetic field lines (gray lines in the top panel) are distributed along this oval every $1^\circ$ in latitude and every 00$:$20 in LT. Each field line is populated by individual radio sources with frequencies covering the SKR spectrum, from 10 to 800~kHz, at the RPWS highest spectral resolution. Each radio source radiates within a cone of emission, whose aperture $\theta_{obs}$(f) is taken from the black line in Figure \ref{fig7}, with a thickness of $\Delta\theta = 5^\circ$. Only a fraction of all simulated radio sources (black crosses) is visible from Cassini (blue crosses). The black dashed lines indicates the position of the guide rotating meridian, the green solid line indicates LT$_{sc}$ and the red meridian indicates noon.}
\label{figsup2}
\end{figure*}

Figure \ref{figsup2} shows the simulated sources as seen from the spacecraft (top panels) and as projections along magnetic field lines onto the polar plane (bottom panels) in the conditions prevailing at times of the four SKR predicted/observed maxima (the vertical white dashed lines in Figure \ref{fig1}a). By definition, the corresponding rotating guide meridian thus lies at 08$:$00~LT. The source region is chosen to magnetically map to a model circular oval (seen in bottom panels). While we used the real position of the UV oval in a previous study \citep{Lamy_these_08}, we could not do similarly here because UV observations were not available over the whole DOY 27-28 interval. Only a fraction of the simulated sources is visible from Cassini (blue crosses). SKR sources at the highest frequencies (closest to the planet, at lowest altitudes) are visible close to the sub-spacecraft meridian, while sources at decreasing frequencies (located farther from the planet, at higher altitudes) are visible at increasing relative longitudes. The SKR visibility pattern evolves with time with the source-observer geometry. In Figure \ref{figsup2}a, visible sources belong to two separate groups of field lines defined by LT$_{sc}\pm$02$:$30 to 04$:$00~LT. This result compares with the westward visibility window identified above or observed in past studies \citep{Cecconi_JGR_09}. In Figure \ref{figsup2}d, visible sources belong to a single, longitudinally continuous, group of field lines defined by LT$_{sc}\pm$03$:$00. 

Figure \ref{figsup}d goes farther by simulating SKR sources during the brightening discussed in section \ref{plasmoid} (close to the third SKR peak), but with a realistic source region mapping to the main UV oval directly fitted from image e$_1$. Here, visible sources were located along a single, longitudinally continuous group of field lines, the west and east branches joining at high frequencies (low altitudes). This slightly differs from Figure \ref{figsup2}c and illustrates the influence of the source region on the SKR visibility.

As an attempt to quantify the fraction of S visible sources and the fraction of S visible power for simulations displayed in Figures \ref{figsup},\ref{figsup2}, we performed a parametric study whose results are given in table \ref{tab_skr}. To do so, we additionally considered a realistic intensity, given by the median SKR spectrum \citep{Lamy_JGR_08a} varying with LT as observed \citep{Lamy_JGR_09}. RPWS detected 3.6 to 4.8\% of all sources and 3.4 to 11\% of the total power. The fraction of visible sources and power first evolved with time with the source-observer geometry, with increases by $25\%$ and $70\%$ respectively from the first to the last SKR burst. These values did not significantly evolve by changing the SKR spectral range (second line). They slightly changed by changing the beaming angle (fourth line), especially for the visible power which decreased to 3.7 to 4.0\%. Finally, and as expected, they rapidly dropped (by a factor of $\sim$~5 for the fraction of visible sources) when the thickness of the beaming angle reduces to $\Delta\theta=1^\circ$ (third line). 
In addition, these values need to be further decreased by a factor of 2 if we include northern sources, which were not visible for most of the interval.

\begin{table*}[!h]
\center
\begin{tabular}{c|c|c|c|c|c}
  Time&DOY 27 05$:$25&DOY 27 16$:$10&DOY 28 03$:$00&DOY 28 03$:$10&DOY 28 13$:$45\\
  \hline
  Source region&Circular oval&Circular oval&Circular oval&UV oval e$_1$&Circular oval\\
  \hline
  \hline
  Visible sources ($\theta_{obs}$(f),$\Delta\theta=5^\circ$,10-800~kHz)&3.6\%&3.5\%&4.1\%&4.6\%&4.8\%\\
  Visible power ($\theta_{obs}$(f),$\Delta\theta=5^\circ$,10-800~kHz)&3.4\%&4.1\%&6.7\%&8.5\%&11\%\\
  \hline
  \hline
  Visible sources ($\theta_{obs}$(f),$\Delta\theta=5^\circ$,20-500~kHz)&3.5\%&3.4\%&3.7\%&4.1\%&4.1\%\\
  Visible power ($\theta_{obs}$(f),$\Delta\theta=5^\circ$,20-500~kHz)&3.3\%&4.0\%&6.4\%&8.2\%&10\%\\
  \hline
  \hline
  Visible sources ($\theta_{obs}$(f),$\Delta\theta=1^\circ$,10-800~kHz)&0.73\%&0.57\%&0.93\%&0.82\%&1.1\%\\
  Visible power ($\theta_{obs}$(f),$\Delta\theta=1^\circ$,10-800~kHz)&0.72\%&0.18\%&2.0\%&1.1\%&2.2\%\\  
  \hline
  \hline
  Visible sources ($\theta_{mod}$(f),$\Delta\theta=5^\circ$,10-800~kHz)&3.7\%&3.6\%&3.5\%&4.1\%&4.0\%\\
  Visible power ($\theta_{mod}$(f),$\Delta\theta=5^\circ$,10-800~kHz)&2.1\%&2.4\%&2.6\%&4.0\%&5.0\%\\  
\end{tabular}
\caption{Percentage of S SKR visible sources and S SKR visible power at times matching the simulations of Figure \ref{figsup2} (column 2,3,4 and 6) and Figure \ref{figsup}d (column 5). $\theta_{obs}$(f) ($\theta_{mod}$(f)) refers to the observed (modeled) beaming angle plotted in black (gray) in Figure \ref{fig7}, and $\Delta\theta$ to its angular thickness.}
\label{tab_skr}
\end{table*}

\subsection{Dynamic spectra}
\label{dynamic_spectra}

Dynamic spectra can be simply built over the whole interval DOY 28-27 by identifying for each time step the frequency and the intensity of visible sources. In the previous section, we showed that an instantaneous observation of a broadband spectrum simultaneously scans radio sources generally located along two separate groups of field lines, symmetrically distributed on both sides of sub-Cassini meridian. Conversely, a single active field line in corotation, successively crossing these two regions, yields vertex-early ($\le$~LT$_{sc}$) and vertex-late ($\ge$~LT$_{sc}$) emissions forming a radio arc in dynamic spectra. When the visible region is continuous in longitude (as in Figures \ref{figsup}d,\ref{figsup2}d), the vertex-early and vertex-late portions of the arc will join at highest frequencies.

\begin{figure*}
\centering\includegraphics[width=1.0\textwidth]{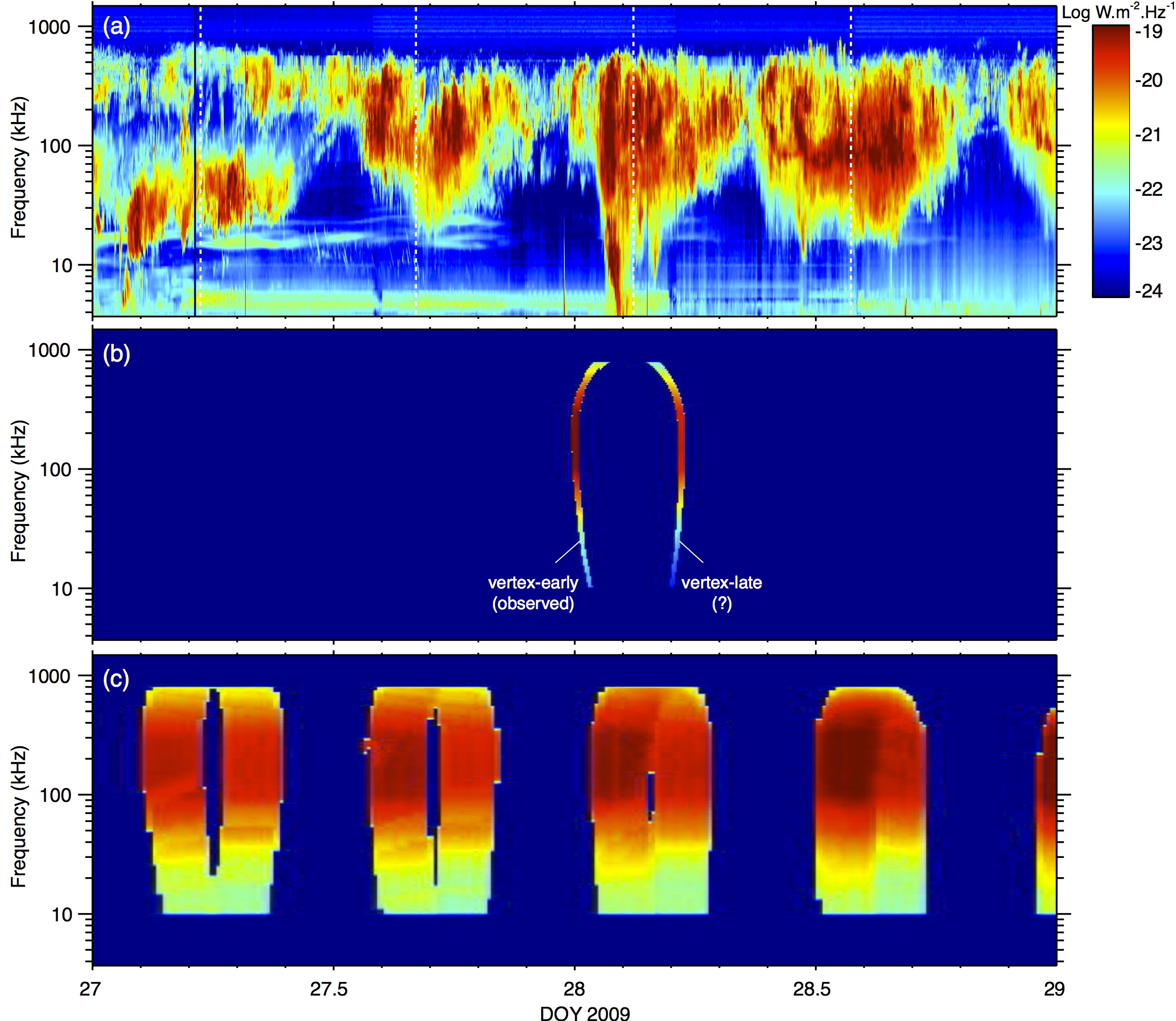}
\caption{SKR observed (a) and modeled (b,c) dynamic spectra with simulation parameters already used in Figure \ref{figsup2} and different source regions. Panel (b) models an active magnetic field line connected to the sub-corotating auroral spot Sp of Figures \ref{fig1}c,\ref{fig2}b. As expected, the vertex-early portion of the modeled arc (center of curvature on the right) reproduces the observed one. Panel (c) models an active auroral source region mapping to the circular oval of Figure \ref{figsup2}, $90^\circ$ wide in longitude and rotating at the S SKR phase. The simulation nicely reproduces the broad observed rotational modulation, and the evolution of the SKR bursts with time.}
\label{fig8}
\end{figure*}

Figure \ref{fig8}b models the dynamic spectrum produced by an active magnetic field line connected to the sub-corotating auroral spot Sp of Figures \ref{fig1}c,\ref{fig2}b. As expected, the vertex-early part of the modeled arc reproduces the observed one. This illustrates how multiple isolated auroral features can account for the observed SKR dynamics at timescales ranging from minutes to hours (several other examples of SKR arcs are observed before DOY 27.5). The vertex-late branch is not observed, possibly because the spot had vanished before it reached the eastward part of the visibility window (indicated by the bottom pair of green lines in Figure \ref{fig2}b).

Figure \ref{fig8}c models the dynamic spectrum produced by a large scale source region mapping to the circular oval of Figure \ref{figsup2}, $90^\circ$ wide in longitude and rotating at the S SKR phase. These conditions reproduce the active auroral region studied in section \ref{sources}. The simulated dynamic spectrum nicely reproduces the main features of the observed SKR rotational modulation : four broadband bursts peak at the right phase and reveal the search-light nature of the SKR modulation identified by \citet{Lamy_PRE7_11} and \citet{Andrews_JGR_12}. Similar results are obtained from the alternate $\theta(f)$ displayed in Figure \ref{fig7}. An important result is that the temporal width of the burst is primarily controlled by the longitudinal width of the active region. 

Figure \ref{fig8}c then provides further insights to refine our interpretation of Figure \ref{fig8}a. Interestingly, the first two bursts divide in two vertex-early and vertex-late wide arches, with a gap in-between. As the longitudinal width of the simulated active region is fixed, this illustrates the effect of the slightly increasing latitude of Cassini on the radio visibility window. Similarly to individual arcs, the vertex-early portion of each burst refers to sources located westward of Cassini, and conversely for its vertex-late portion. Beyond this property, the modeled SKR bursts remain approximately constant with time, with a slowly varying spectrum, as observed. An exception is the first observed SKR burst, which peaks at low frequencies, from 10 to 70~kHz. The discrepancy with the first simulated burst indicates that the most intense SKR sources are here located farther along the field line than simulated. As the primary condition for CMI (plasma frequency much lower than cyclotron frequency) is permanently fulfilled from 1000~kHz down to a few kHz, the generation of radio emission may be controlled only by the presence of unstable electron distributions, which implies that the acceleration region itself was located farther from the planet along auroral field lines. The intensity and the peak of the emission did not significantly change from the vertex-early arch (dawnside sources) to the vertex-late one (duskside sources) of the first burst, suggesting a stead-state source of acceleration for at least a few hours. In addition, the SKR emission visible above 200~kHz before DOY 27.5, spectrally separated from the low frequency portion of the spectrum, indicates a secondary source region.

\section{Discussion}
\label{discussion}

\subsection{Co-existence of corotational and sub-corotational auroral dynamics}

One of the important findings of this study is the evidence that sub-corotating auroral features coexist with corotating ones, as detected in radio and UV emissions associated with the main oval. On the one hand, this leads to a new understanding of short timescale SKR structures observed in dynamic spectra which will provides better diagnosis capabilities. On the other hand, it seems that a longitudinal active sector of auroral field lines is continuously populated with accelerated particles by a perturbation in corotation, likely the S corotating field-aligned current system identified from in situ magnetic measurements \citep[and refs therein]{Andrews_JGR_10a}, while isolated field lines, once activated, would move together with the ambient plasma. Solving this paradox is beyond the scope of this paper, but we remark that the frozen-field theorem applies in ideal MHD conditions, with a perfectly conducting plasma and no parallel electric field. The latter condition, however, is likely to be wrong along field lines supporting field-aligned currents.

\subsection{Plasmoid occurrence}

This study also suggests regular nightside injections of energetic particles over (at least) four successive rotations, likely internally triggered. Among them, a particularly powerful event may be associated with a significant nightside plasmoid ejection. The activated ENA region connects to enhanced auroral emissions, whose intensity rose with corotation through dawn, in agreement with the LT intensity profile derived from previous statistical analysis. This matches the results obtained from the analysis of a similar sequence of colocated ENA, radio and UV enhancements \citep{Mitchell_PSS_09}, and the conclusion of a statistical study of the occurrence of plasmoid release by in situ magnetic field measurements and remote sensing of SKR \citep{Jackman_JGR_09}. \citet{Mitchell_PSS_09} suggested that reccurrent plasmoid releases can be quasi-periodically driven by the continuous mass loading of the inner magnetosphere by Enceladus subject to the magnetospheric rotation, inspired by steady-state processes inherent to the jovian magnetotail \citep{Vasyliunas_83}. Such a process has been recently modeled for Saturn \citep{Zieger_JGR_10,Jia_JGR_12}. The rapid reconfiguration of the magnetic field on the planetward side of the reconnection site would energize ions up to several hundreds of keV and lead to an enhanced, localized, region of (hot) plasma pressure, able to drift around Saturn and drive a three-dimensional current system. This is the process behind the underlying pressure of the partial ring current at Earth, except that the terrestrial partial ring current is rather stationary around midnight \citep{Ijima_JGR_90}.

\subsection{Polar and equatorward atmospheric emissions}

Previous studies essentially noticed polar emissions in the IR spectral range, and concluded they were a specific behavior of polar H$_3^+$. Here, significant polar emissions have been observed both in the IR and the UV. We measured higher polar-to-main-oval ratios for radiated H Ly-$\alpha$ than for H$_2$, which we suggested to be caused by softer electron precipitations at higher altitude deposition layer in the polar region ($\le10$~keV) with respect to the main oval (up to 20~keV) and/or lower fluxes. 

The polar-to-main-oval ratio of H$_3^+$, whose altitude of emission is close to that of H$_2$, remains larger than both H Ly-$\alpha$ and H$_2$. We established that the temperature is roughly constant across the auroral region, and thus does not control the latitudinal variation of H$_3^+$ emission. Considering fixed temperature and precipitating flux, \citet{Tao_Icarus_11} showed that the IR relative to UV volume emission rate increases with decreasing energy below 10~keV. Softer polar electrons thus appear as a natural cause for a larger H$_3^+$ polar-to-main-oval ratio. Alternately, it could also result from a fainter main IR oval, as \citet{Tao_Icarus_11} showed that the UV relative to IR volume emission rate increases both with a fixed precipitating flux and increasing energies beyond 10~keV or with fixed energies and increasing precipitating fluxes.

Equatorward emissions are faint without a visible LT dependence, suggesting a weakly efficient, steady-state, acceleration mechanism. The latitudinal range found for the emission peak, $-68^\circ$ to $-70^\circ$, yields the equatorial range 6 to 7.3~R$_S$ from the magnetic field model we used (see section \ref{skr_modeling}). This range is significantly beyond the 4~R$_S$ orbital distance ($-62^\circ$ footprint latitude) of Enceladus. It is however in excellent agreement with the 7 to 9~R$_S$ range where two co-existing populations of hot plasma have been observed to peak : one suprathermal ($\sim$~1~keV), roughly isotropic, tenuous component and one warm ($10-100$~eV), denser, field-aligned component \citep{Schippers_JGR_08,Schippers_JGR_12}. Electrons originating from these populations have been proposed by \citet{Grodent_JGR_10} to be responsible for the observed equatorward aurorae. The observed variation in latitude of these emissions, resembling that of the main oval, may in this frame provide a useful diagnosis of the variable location of these hot plasma populations.

\begin{table*}[!h]
\center
\begin{tabular}{c|c|c|c|c}
  Spectral range&UV ($80-165$~nm)&IR ($3-5~\mu$m)&Radio ($1-1000$~kHz)\\
  \hline
  Type of emission&H$_2$ and H Ly-$\alpha$ (H Ly-$\alpha$ alone)&H$_3^+$&SKR\\
  \hline
  Radiated power&$1.3-4.3\times10^{11}$~W ($0.7-5.6\times10^{10}$~W)&$2-4\times10^{11}$~W&$0.1-2\times10^{9}$~W\\
  \hline
  Emission efficiency&21\% (3\%)&20-40\%&$\ge0.1\%$\\
\end{tabular}
\caption{Total power radiated by kronian auroral emissions and median efficiency with respect to the total input power (varying within the $7.4\times10^{11}$ to $2.1\times10^{12}$~W range, with a median value of $8.6\times10^{11}$~W), as derived in section \ref{energy}.}
\label{tab_fin}
\end{table*}

\subsection{SKR sub-components?}

Simulations of section \ref{skr_modeling} showed the strong correspondence between the main atmospheric aurorae and the bulk of SKR. We can check if any SKR sub-component can be associated with either polar and/or equatorward emissions, assuming the presence of electron distributions unstable to CMI on the corresponding field lines. Equatorward emissions are faint, fixed in LT and located at the footprint of inner magnetic field lines. The theoretical spectral range for radio emission, over which the primary CMI condition needs to be fulfilled, is necessarily reduced at both low and high frequencies. As a result, any radio counterpart of equatorward emissions would be embedded within the dominant SKR signal, without any easy means to discriminate it either by its intensity and/or by its dynamics. The case of polar emissions is more favorable, since they are quite bright relative to the main oval. Their higher latitude implies a spectral range for CMI-driven emission that theoretically overcomes that of the main SKR. A simulation of a steady-state polar arc over $-76$ to $-78^\circ$ latitude and between 06$:$00 and 15$:$00 (not shown) suggests that its dawnside and duskside parts shall be continuously visible from Cassini. However, as they do not (sub)co-rotate, they would not produce arc-shaped features in the time-frequency plane. We can thus only expect short-lived radio bursts simultaneous to polar emissions observed in UV/IR images. The SKR dynamic spectrum does not reveal obvious signs of activity correlated with sporadic polar aurorae. Therefore, the existence of kilometric radiations associated with equatorward and polar auroral emissions, if any, remains an open question.

\section{Summary}

We have analyzed two days of combined Cassini measurements of Saturn's aurorae at radio, UV and IR wavelengths and of ENA emissions, in an attempt to give a comprehensive view of the nature and the dynamics of kronian auroral processes over the timescale of a planetary rotation.

We identified three distinct source regions, seen in both UV and IR domains : (i) a circumpolar quasi-circular main oval between $-70^\circ$ and $-72^\circ$ latitude, associated with the bulk of SKR emission, (ii) polar emissions intermittently observed along (part of) a ring of emission extending between 06$:$00 and 16$:$00 from $-75^\circ$ to $-82^\circ$ latitude and (iii) a faint arc of emission fixed in LT between $-68^\circ$ and $-70^\circ$ latitude and extending from dawn to dusk through noon.

The main oval is subject to a clear rotational modulation. It contains a $50-90^\circ$ wide active region corotating at the S SKR phase, which triggers powerful SKR bursts each time it passes through the dawn sector, and isolated features sub-corotating at velocities consistent with that of the ambient equatorial magnetospheric cold plasma. The main bright region was particularly enhanced after a powerful nightside injection of energetic ions at the S SKR phase. The intensity of the equatorward auroral emissions was almost uniform and fixed in LT, likely associated with the suprathermal electron population between 7 and 9~R$_S$. In contrast, sporadic polar emissions, varying at timescales down to 15~min, were associated with open field lines and are expected to be under solar-wind controlled processes.

We were able to determine the energy of primary precipitating electrons from FUV spectroscopy, by the determination of the selective absorption of the short end of the spectrum. The calculated energies remain within the 10$-$20~keV range all over the auroral features. Below this lower limit, absorption becomes undetectable and this technique is ineffective. We found that the measurement of the ratio of H Ly-$\alpha$ to H$_2$ brightnesses gives an alternate and sensitive diagnosis for electrons of lower energies. Overall, brightnesses of 10~kR and electron energies of 10~keV are typical along the main oval. The upper limit of 20~keV matches active regions of the oval such as the bright rotating region. The large variations of the radiated fluxes are mostly attributed to the increase of the electron flux rather than of their energy. Polar emissions are fainter, and high H Ly-$\alpha$ to H$_2$ brightness ratios in this region suggest electrons below 10~keV. Equatorward emissions are even fainter. 

IR spectroscopy provided estimates of the thermospheric temperatures, found to be roughly constant over the whole auroral region and with time, around $\sim$410~K. On average, the polar-to-main-oval intensity ratio of H$_3^+$ is larger than that of H Ly-$\alpha$, itself larger than that of H$_2$. This discrepancy was attributed to differences in energies and/or fluxes of precipitating electrons rather than to temperature variations. 

The radiated powers were estimated separately, with an excellent temporal correlation between radio and UV emissions (the limited SNR of most H$_3^+$ images prevented a search for a similar degree of association). The budget of auroral energy dissipated through the various radiating processes is summarized in table \ref{tab_fin}.

Finally, we were able to develop further the simulation of the observed SKR features. We successfully modeled the rotational modulation, definitely confirming its search-light nature, and we showed that most of the SKR sub-structures, varying at timescales from minutes to hours, are radiated along field lines connected to isolated sub-corotating hot spots within the main auroral oval. We finally demonstrated that only a few percent of the total SKR power was visible from the spacecraft during the investigated interval, and the SKR visibility changes significantly with the source-observer geometry.

\appendix

\section{UVIS data processing}
\label{uvis_processing}

Within the spectral range of the UVIS-FUV channel, H Ly-$\alpha$ and H$_2$ auroral emissions are superimposed on top of a non negligible background, while the Extreme UV (EUV) part of the H$_2$ spectrum is not covered. To solve these issues and derive meaningful physical units, we applied the following additional processing to UVIS data, illustrated with the example of one image in Figure \ref{fig_app4}.

\subsection{Background correction}

\begin{figure}
\centering\includegraphics[width=0.5\textwidth]{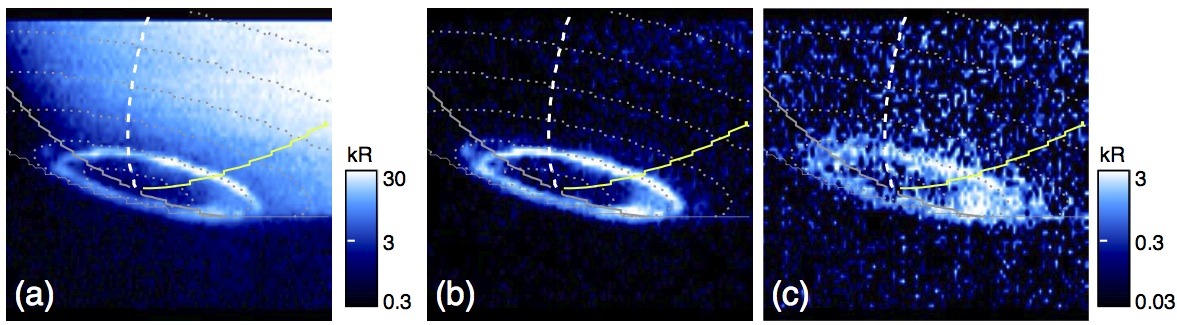}
\caption{(a) Original UVIS image $a_1$, integrated from 120 to 190~nm. (b) Processed image of total H$_2$ emission from 80 to 170~nm, corrected for background emission and for absorption as detailed in the main text. (c) Same as (b) for H Ly-$\alpha$. The left color bar refers to panels (a) and (b). The right color bar refers to panel (c). Grids of coordinates are replicated from animation S1.}
\label{fig_app4}
\end{figure}

The extraction of purely auroral H and H$_2$ emissions, primarily requires a specific background subtraction. The latter includes reflected sunlight, reflected stellar light, H$_2$ atmospheric airglow, stars crossing the field-of-view, H Ly-$\alpha$ diffusion and instrumental effects (see for instance \citep[and refs therein]{Gustin_Icarus_10}), present in Figure \ref{fig_app4}a. Taking advantage of the knowledge of the FUV spectrum every spatial pixel (in kR.nm$^{-1}$ after calibration), we applied a separate background correction separately for pixels belonging to three distinct regions : the sunlit and the nightside parts of the planetary disc, and the sky beyond the limb.

\begin{figure}
\centering\includegraphics[width=0.5\textwidth]{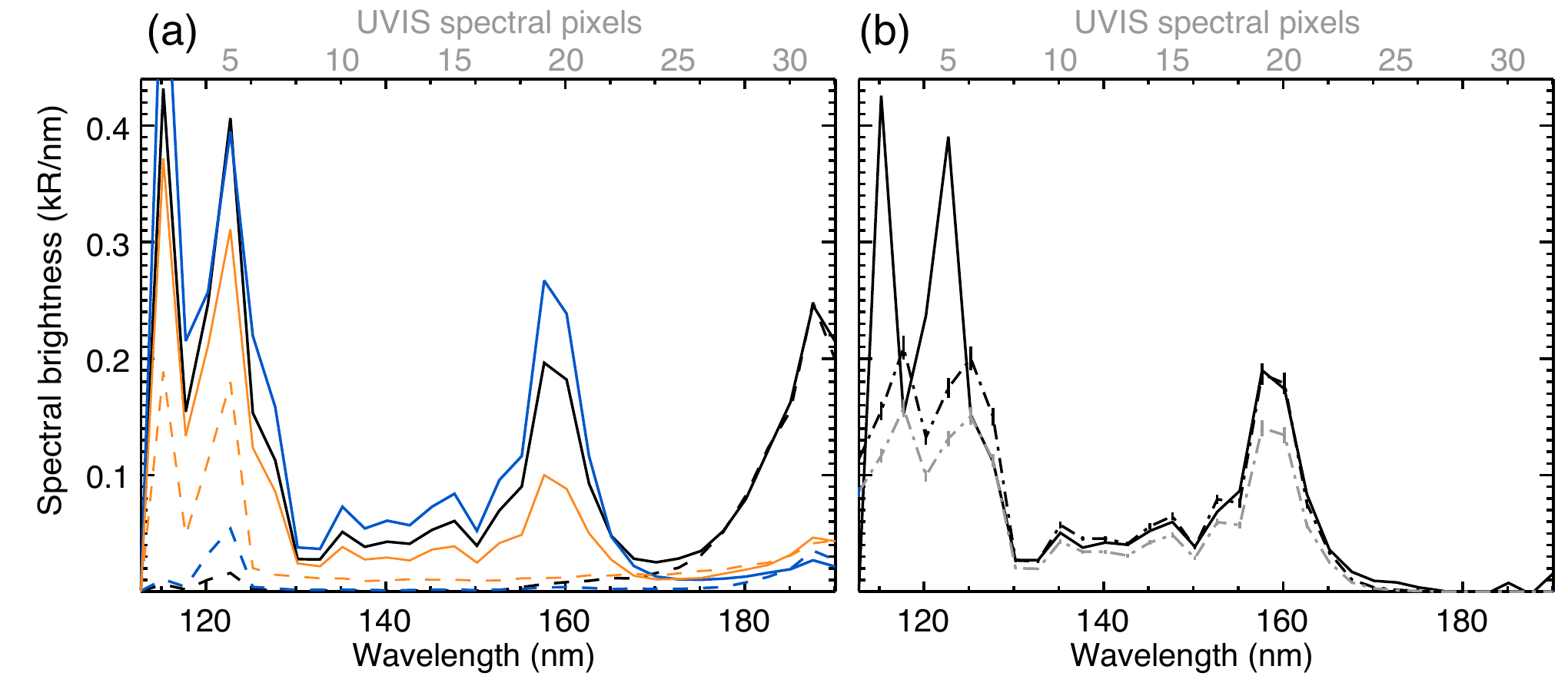}
\caption{UVIS FUV spectra as a function of wavelength (bottom axis) or spectral pixels (top axis). (a) Auroral (solid) and background (dashed) average spectra derived separately over the sunlit disc (black), the nightside disc (blue) and the sky (orange). Auroral spectra were constructed from pixels with brightnesses $\ge1$~kR of total H$_2$, restricted to high latitudes ($\le-65^\circ$) for the disc. Conversely, background spectra were built from low latitude pixels for the disc or pixels at the lower edge of the slit (spatial pixels 4 to 19) for the sky, normalized to auroral spectra over the 180-190~nm range. (b) Average auroral spectrum of the sunlit disc corrected for background (solid). Synthetic spectra of H$_2$ degraded at the UVIS resolution (dotted-dashed), once normalized over 123-130~nm (gray, subject to methane absorption along the line-of-sight) and 144-166~nm (black, unabsorbed). Vertical error bars give $\pm5\%$ levels, taken as a typical uncertainty of the fit. The ratio between absorbed and unabsorbed synthetic spectra yields a non negligible absorption of $1.33\pm0.13$.}
\label{fig_app1}
\end{figure}

Average spectra of sunlit and nightside portions of the disc were built separately from non-auroral pixels only, assuming that individual spectra are similar in shape and that they only vary with LT and latitude by their magnitude. Resulting spectra are displayed in Figure \ref{fig_app1}a by black and blue dashed lines respectively. Reflected sunlight at long wavelengths is the dominant source of background for the sunlit disc, while its nightside portion (between the terminator and the limb) is primarily affected by H Ly-$\alpha$ emission. Each individual spectrum from the planetary disc was individually corrected for its corresponding background (sunlit or nightside) normalized over the 180-190~nm range, free of H$_2$ bands.

Outside the planetary limb, the background is dominated by stellar light, H Ly-$\alpha$ diffusion and significant instrumental leakage along the slit for bright emissions. As a result, the sky brightness measured over the (southward) lower edge of the slit varies with time, as a function of the intensity recorded by the rest of the slit. In addition, low levels of signal measured beyond the limb prevent a reliable fit of an average sky background (orange dashed line in Figure \ref{fig_app1}a) to individual spectra as above. To correct for this time-variable contribution, we thus processed each 8~s long slit measurement separately, for each of which a mean sky spectrum built from non-auroral pixels of the lower edge of the slit was subtracted from all the spectra from spatial pixels beyond the limb.

\subsection{Auroral brightness corrected for absorption}

\subsubsection{Auroral emission of H$_2$}

H$_2$ emissions consists of a series of lines within the 80 to 170~nm range (transitions from the B, C, B$'$, D, B$''$, D$'$ states to the ground state). The Lyman (B) and Werner (C) bands together with the Lyman continuum prevail in the FUV domain, while representing 90\% of the total UV brightness of H$_2$ \citep{Menager_these_11,Gustin_JGR_12}. To retrieve the total auroral emission radiated by H$_2$ over the whole 80-170~nm range, we used an H$_2$ synthetic spectrum (D. Shemansky, personal communication) accounting for all the transitions.

Once degraded at the UVIS spectral resolution, this synthetic spectrum can be fitted to individual spectra over the 144-166~nm range (black dotted-dashed line in Figure \ref{fig_app1}b), where methane absorption is negligible. The total auroral emission radiated by H$_2$ (excluding H Ly-$\alpha$) from 80 to 170~nm and corrected for absorption is then reconstructed using the measured spectrum above 144~nm and the normalized synthetic one below. This processing is illustrated in Figure \ref{fig_app4}b.

The Rayleigh (1~R~=~10$^6$ photons.s$^{-1}$.cm$^{-2}$ in 4$\pi$ sr), originally introduced by aeronomists for investigating atmospheric photochemistry \citep{Baker_AO_74}, is a brightness unit commonly used with UV aurorae. Although we are primarily interested in the energy (rather than photon) radiated flux for power budget considerations, we chose total kR of H$_2$ as a convenient unit for UV images shown in this study. Indeed, by reducing this brightness by 10\% to restrict to emissions of the H$_2$ continuum, Lyman and Werner bands only, we could directly estimate the power input carried by precipitating electrons through the standard conversion factor of $\sim$~10~kR radiated per incident mW.m$^{-2}$ \citep{Gerard_JGR_82,Waite_JGR_83}.

\subsubsection{Auroral emission of H Ly-$\alpha$}
\label{appendix_absorption}

It is more difficult to retrieve the auroral H Ly-$\alpha$ contribution, which concerns a reduced bandwidth, covered by two spectral pixels from 119 to 123~nm at the UVIS resolution, and which superimposes on top of H$_2$ emission and is subject to methane absorption.

The method used to retrieve the H Ly-$\alpha$ emission radiated before absorption is illustrated with the auroral average spectrum of Figure \ref{fig_app1}b, derived from sunlit auroral pixels whose H$_2$ brightness exceeds 1kR (more than 12000 individual spectra). This spectrum was fitted by a synthetic H$_2$ spectrum over 123-130~nm (gray dotted-dashed) to determine the residual contribution of auroral H$_2$ close to the Ly-$\alpha$ line. The uncertainty on the fit is small, with vertical error bars giving $\pm$5\% levels as an estimate. Once the H$_2$ contribution was subtracted, the single H Ly-$\alpha$ emission was eventually corrected for absorption, given by the ratio between unabsorbed and absorbed H$_2$ spectra, which reaches $1.33\pm0.14$ for this spectrum. Values larger than unity indicate the presence of absorption. This absorption factor is equivalent to normalized color ratios (CR), various definitions of which have been used in the literature. The absorption factors found for the other two auroral average spectra displayed in Figure \ref{fig_app1}a (blue and orange solid lines) are $1.28\pm0.13$ and $0.95\pm0.10$ respectively. The latter value indicates that emissions from outside of the planet were not absorbed on average. An example of the H Ly-$\alpha$ image resulting from this processing is displayed in Figure \ref{fig_app4}c.

In terms of intensity radiated by auroral sources, the sorting between sunlit, nightside and sky regions is irrelevant. We thus now consider all the auroral emissions together in a single average spectrum. The corresponding absorption remains significant at $1.26\pm0.13$. As shown by Figure \ref{fig_ratios} and table \ref{tab1}, the final brightness at H Ly-$\alpha$ (integrated over the 120-123~nm range) corrected for absorption reaches 12\% of the total H$_2$ brightness, while its contribution to the total signal between 120 and 123~nm reaches 58\%. Therefore, while we chose to derive H Ly-$\alpha$ corrected for absorption for all UVIS pixels separately, an alternate and simple estimate of the total auroral brightness of H$_2$ and H Ly-$\alpha$ can be obtained by adding a 13\% mean contribution to that of H$_2$.

In the above discussion, we assumed electrons as the predominant sources for UV emissions. Protons may represent a secondary source of excitation (unreported so far at Saturn to our knowledge) and their presence can in principle be checked from a careful spectral analysis \citep{Rego_JGR_94}. This is left for further studies.

\subsubsection{Uncertainty}

The accuracy of UVIS measurements is limited by random and systematic errors arising from the different processing steps.

Concerning random uncertainties, the UVIS calibration brings to the photon flux of each spatial pixel an intrinsic 15\% error, while the error arising from the background subtraction step is estimated to 5\%. The photon noise corresponding to high SNR measurements of auroral pixels is neglected. These independent contributions yield a quadratic uncertainty of 16\%. The signal is also subject to systematic uncertainties due to the spectral processing. The uncertainty of the fitting step has been estimated to 7\% by comparing the difference in final H$_2$ brightnesses obtained by fitting either the 144-166~nm range or the 155-163~nm one (corresponding to the most intense H$_2$ bands). A last source of uncertainty arises from the choice of the reference spectrum itself, and we checked our extrapolation shortward of 144~nm with alternate H$_2$ synthetic spectra \citep{Menager_these_11}, yielding differences of about 10-15\%. The resulting systematic error reaches about 15\%. 

The uncertainty on H Ly-$\alpha$ is larger, as lower SNR measurements increase the relative contribution of the photon noise. However, the H Ly-$\alpha$ contribution to the total H$_2$ brightness is small (12\% on average).

Overall, the total uncertainty on the H$_2$+H Ly-$\alpha$ brightness is estimated to 30\% for each individual spatial pixel.

\subsection{Total auroral power}

To derive the total auroral power radiated at both H Ly-$\alpha$ and in the H$_2$ bands, we need to consider energy fluxes instead of photon fluxes. The required spectral integration does not affect the contribution of H Ly-$\alpha$ to H$_2$ emission, but increases systematic uncertainties to 20\%. Therefore, the uncertainty of the total auroral power corrected for absorption reaches 35\% for each individual spatial pixel.

However, when considering the auroral power radiated by the whole auroral region, we have to integrate the signal obtained over a large number of pixels, so that random uncertainties rapidly become negligible and the residual final uncertainty on the total auroral power remains about 20\%.

\section{VIMS data processing}
\label{vims_processing}

\subsection{Background correction}

\begin{figure}
\centering\includegraphics[width=0.5\textwidth]{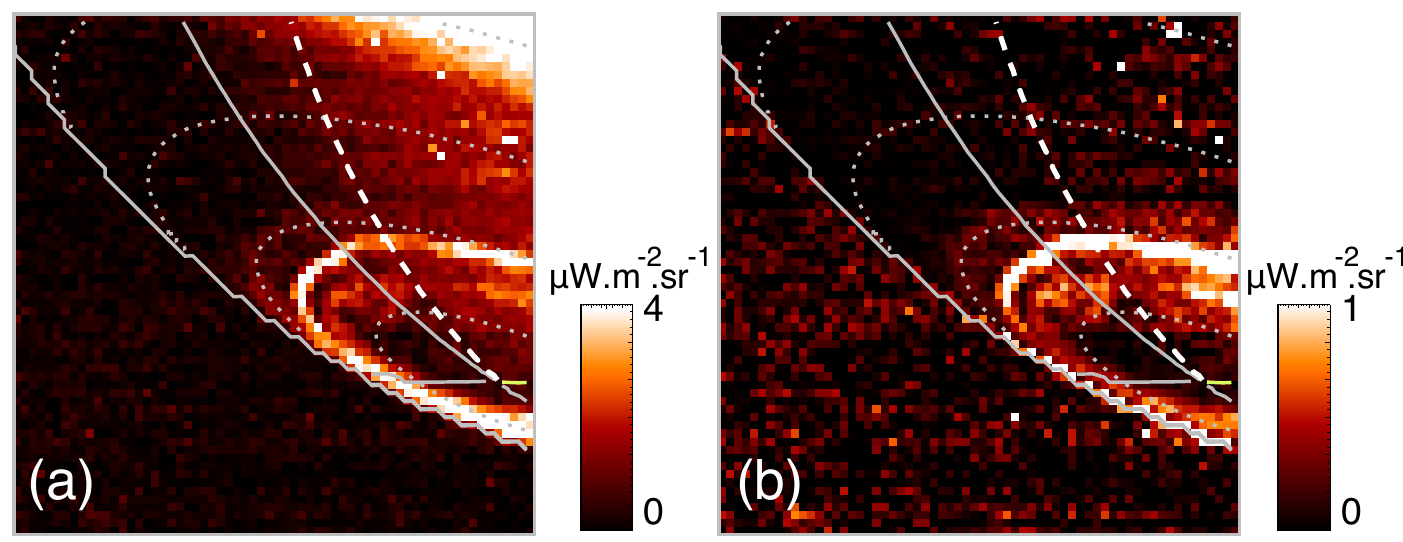}
\caption{(a) Original VIMS image 3, composed of the sum of spectral bins 153, 155, 160, 165 and 168. (b) Processed image, corrected for bad pixels, for background emission and for limb-brightening as detailed in the main text. Grids of coordinates are replicated from animation S1.}
\label{fig_app5}
\end{figure}

The VIMS spectral range covers a large set of H$_3^+$ ro-vibrational lines \citep{Neale_ApJ_96}. For simplicity, images were constructed by integrating 5 spectral bins (153, 155, 160, 165 and 168, centered on 3.417, 3.451, 3.532,  3.616 and 3.667~$\mu$m respectively), where auroral H$_3^+$ is clearly detected. Compared to similar past studies \citep{Stallard_Nature_08b,Melin_GRL_11,Badman_Icarus_11}, more wavelength bins (spectral lines) were considered here to increase the signal-to-noise ratio of short exposure time images. The original calibration pipeline gives data expressed in units of spectral radiance ergs.cm$^{-2}$.s$^{-1}$.$\mu$m$^{-1}$.sr$^{-1}$, here transposed into $\mu$W.m$^{-2}$.sr$^{-1}$ (rather than in Rayleigh) to facilitate comparisons with previous studies.

An example of VIMS image is displayed in Figure \ref{fig_app5}a, where auroral H$_3^+$ emissions are seen on top of a significant background emission. This includes contributions from cosmic rays (hot spots), reflected sunlight, ambient noise level, and sporadic residual instrumental contamination (offseting certain rows of pixels). To correct for this contamination, the following pipeline was applied to each image. Bad pixels were first set to the ambient background. An empirical background model was then built from the sum of 4 spectral bins (150, 162, 166 and 170, centered on 3.368, 3.565, 3.633 and 3.702~$\mu$m respectively), weightened by empirical factors (90, 80, 55 and 70\% respectively). We also checked an alternate background model, inspired from that used by \citet{Badman_Icarus_11}, which provided similar results. Once this model of background subtracted, a simple line-of-sight correction was finally applied to each pixel by multiplying its intensity value by the cosinus of the emission angle \citep{Badman_Icarus_11}. As pixels coordinates are only known within the disc, some residual limb-brightening remain for auroral emission beyond the limb. However, this residual contamination concerns a small amount of pixels, and does not concern polar projections. The result of the above processing step is illustrated in Figure \ref{fig_app5}b.

\subsection{Uncertainty}

The accuracy of VIMS measurements is mostly limited by random errors.

The VIMS calibration is responsible for an intrinsic 15\% error on fluxes (R. Brown, personal communication). We estimated the uncertainty resulting from the background subtraction to $\sim$~15\%. These independent contributions yield a quadratic uncertainty of 21\% for each spatial pixel. This estimate provides a lower limit, as short integration times of most VIMS images are expected to yield a large photon noise.

\subsection{Temperature determination}

Ionospheric temperatures were derived as described by \citet{Melin_GRL_11}. For each VIMS spatial pixel, we first calculated the ratio $b$ given by the sum of spectral bins 153, 155 and 160, corrected for a background model given by the weighted sum of spectral bins 150 and 162, over that of spectral bins 165 and 168, corrected for a background model given by the weighted sum of spectral bins 166 and 170. The temperature was then calculated with the expression : 

\begin{equation}
T(b) = 122 + 289b - 62b^2 + 25b^3
\label{eq_temp}
\end{equation}

The accuracy of this determination depends on the intensity of both the numerator and the denominator of the ratio $b$, themselves dependent on the H$_3^+$ emission intensity. To remove pixels with low SNR, we used a threshold of 0.5~$\mu$W.m$^{-2}$.sr$^{-1}$ on both the numerator and the denominator of $b$. This value was chosen as a typical upper limit for the noise on the numerator and the denominator for spatial pixels beyond the limb.

\begin{figure*}
\centering\includegraphics[width=1\textwidth]{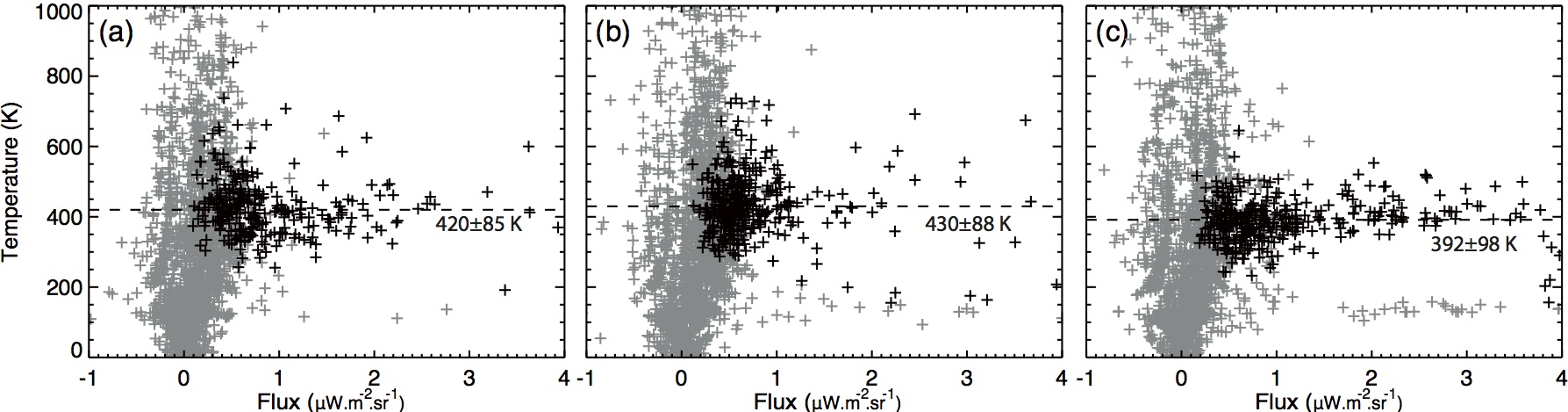}
\caption{Ionospheric temperature as a function of the H$_3^+$ flux (sum of spectral bins 153, 155, 160, 165 and 168) for long exposure time VIMS images 3, 20 and 37 (panels a, b and c respectively). Gray crosses display all spatial pixels. Black crosses restrict to those for which both the numerator and the denominator of $b$ overcome 0.5~$\mu$W.m$^{-2}$.sr$^{-1}$. Black numbers give the mean temperature and standard deviation calculated from black crosses.}
\label{fig_app2}
\end{figure*}

Figure \ref{fig_app2} displays the ionospheric temperature as a function of the corresponding H$_3^+$ flux for the three long exposure time VIMS images 3, 20 and 37. The gray crosses provide the temperature estimates for all pixels. Out of these, pixels with low H$_3^+$ fluxes result in noisy temperatures, whose uncertainty can reach several hundreds of K. The black crosses show the data selection resulting from the above thresholding. The removal of low SNR pixels yields much less dispersed temperatures with median values around 400~K and standard deviations less than 100~K. In addition, we note the absence of any significant correlation between temperatures and H$_3^+$ fluxes. Black crosses (brighter pixels) only were used for the temperatures plotted in Figure \ref{fig6}.

\subsection{Radiated power}

\begin{figure}
\centering\includegraphics[width=.48\textwidth]{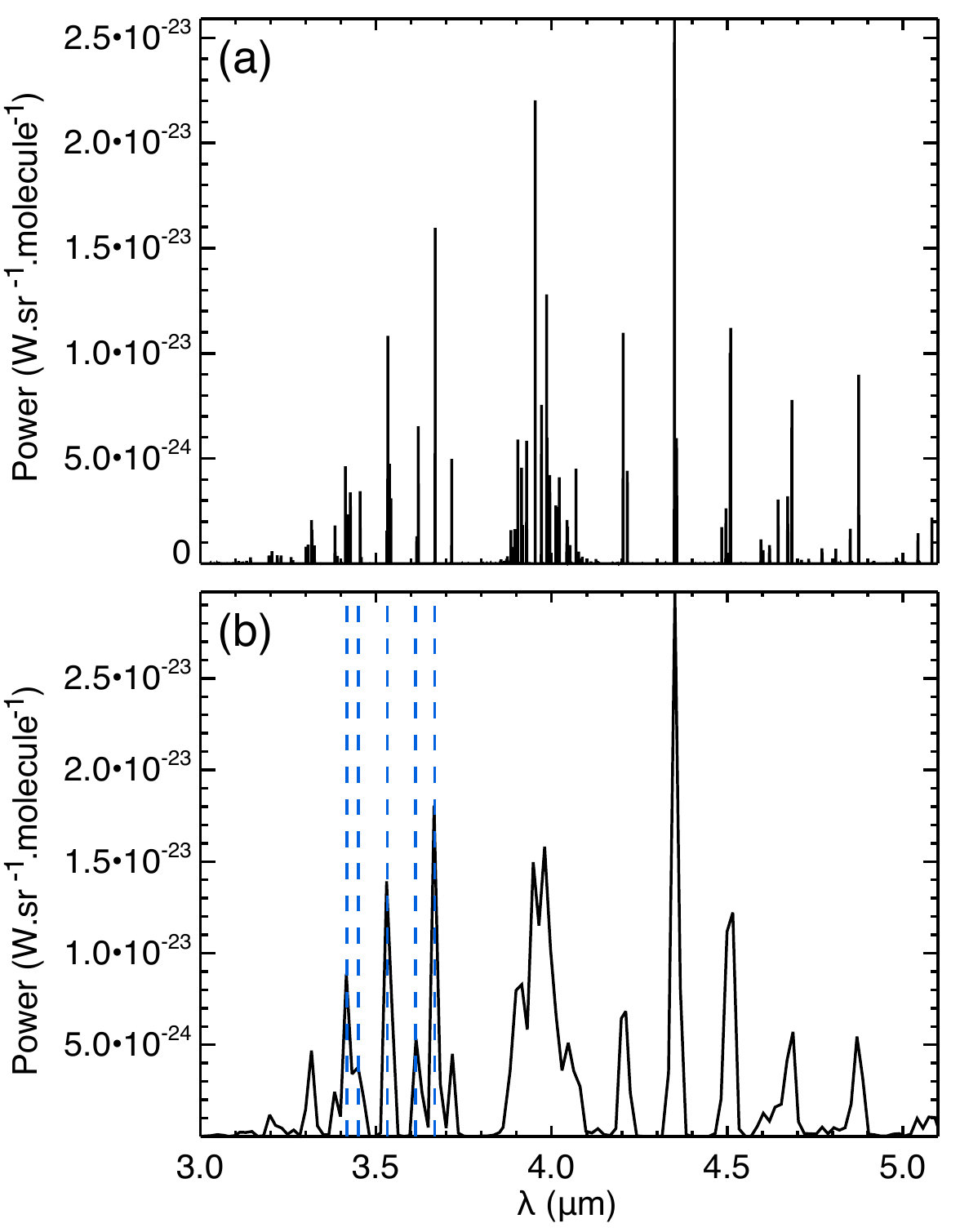}
\caption{(a) Theoretical spectrum of transitions E$_i$ of a single H$_3^+$ molecule at 440~K, calculated from the list of H$_3^+$ lines of \citet{Neale_ApJ_96} and the corrected partition function of \citet{Miller_FD_10}. (b) Same spectrum as simulated \citep{Melin_these_06} through the VIMS instrument. Dashed lines indicate the central wavelengths of VIMS bins used in this study.}
\label{fig_app3}
\end{figure}

Under LTE conditions, the power P$_i$ emitted by a single H$_3^+$ ro-vibrational line can be expressed as :

\begin{equation}
P_i(\omega_{if},T) = \frac{g_f(2J_f+1) \times hc\omega_{if} \times A_{if}}{4\pi Z(T)}\exp^{\frac{-hcE_f}{kT}}
\end{equation}

where $g_f$ is the nuclear spin degeneracy, $J_f$ the angular momentum, $\omega_{if}$ and $A_{if}$ the angular frequency and Einstein coefficient of the transition, E$_f$ the energy of the upper state, $h$ and $k$ the constants of Planck and Boltzmann, $c$ the speed of light and Z(T) the partition function \citep{Miller_FD_10}. P$_i$ depends on the frequency of the transition and the temperature T. For our purpose, we used the list of H$_3^+$ lines produced by \citet{Neale_ApJ_96}. We selected ortho/para transitions between 1.5 and 6.7~$\mu$m, in the same way as \citet{Melin_these_06}, but without any selection on Einstein coefficients, which yields a total of 15210 individual lines. We also used the corrected partition function proposed by \citet{Miller_FD_10}, who showed that the partition function used by \citet{Neale_ApJ_96} underestimates P$_i$ (by about 30\% at T~=~500~K). Figure \ref{fig_app3}a displays the theoretical spectrum of a single H$_3^+$ molecule at 440~K. The power P$_{mol}$(T) radiated by a single H$_3^+$ molecule directly derives from the spectral integration of P$_i$(T). 

The total H$_3^+$ power P$_{tot, pix}$ radiated by each spatial pixel is then written as :

\begin{equation}
P_{tot, pix}(T) = N_{pix}\times P_{mol}(T)\times4\pi\times a_{pix}
\end{equation}

where N$_{pix}$ is the H$_3^+$ column density corresponding to the pixel, a$_{pix}$ its projected area on the observing plane and 4$\pi$ the solid angle normalization factor. This normalization is identical to that applied to the UV power. It is sometimes restricted to $2\pi$ in the literature, in which case the calculated quantity is referred to as the cooling, and deals with the fraction of the emission radiated spaceward only. The calculation of P$_{tot, pix}$ thus requires the knowledge of the temperature T, the column density N$_{pix}$ and the projected area for each VIMS pixel. 

Temperatures were determined as described in section \ref{temperatures}. Although VIMS measurements are noisy, the temperature is found to be roughly constant over the auroral region, so that we chose a single temperature for all auroral VIMS pixels of a given image. 

N$_{pix}$ can then be derived from the ratio of the VIMS measurement at a given wavelength $\lambda$ (expressed in W.m$^{-2}$.sr$^{-1}$) to the corresponding VIMS simulated response for a single molecule at the same wavelength at temperature T (expressed in W.sr$^{-1}$.molecule$^{-1}$). The VIMS simulated spectrum is plotted for T~=~440~K in Figure \ref{fig_app3}b. To increase the SNR of this determination, we used the H$_3^+$ flux computed for images, obtained by summing up 5 spectral bins.

Finally, the total H$_3^+$ power P$_{tot}$ radiated by the whole S auroral region is obtained by integrating P$_{tot, pix}$ over the auroral region. Since only a portion of it is visible by VIMS, we took advantage of the similarity found between simultaneous FUV and IR auroral features to apply a correction factor $\alpha$, derived from the closest in time UVIS image, to extrapolate P$_{tot}$ to the entire auroral region.

\begin{equation}
P_{tot}(T) = \alpha \sum_{pix} \times P_{mol, pix}(T)
\end{equation}

This calculation was made for long exposure time VIMS images 3, 20 and 37. The results are summarized in table \ref{tab_H3p_power}. The uncertainty on P$_{tot}$ is dominated by that on $\alpha$, which reaches $\sim$~30\%.

\begin{table*}[!h]
\center
\begin{tabular}{c|c|c|c}
  VIMS data&Image 3&Image 20&Image 38\\
  \hline
  Mean temperature T K)&$420\pm85$&$430\pm88$&$392\pm88$\\
  Median column density N (m$^{-2}$)&$1.0\times10^{16}$&$1.3\times10^{16}$&$1.9\times10^{16}$\\
  Pixel projected area a (km$^2$)&($1120\pm10$)$^2$&($1120\pm10$)$^2$&($1150\pm10$)$^2$\\
  Integrated number of particles&$3.8\times10^{31}$&$4.6\times10^{31}$&$1.0\times10^{32}$\\
  Geometric correction factor $\alpha$&$3.7\pm1$&$2.3\pm1$&$2.3\pm1$\\ 
  Total radiated power P$_{tot}$ (W)&$3.7\times10^{11}$&$3.3\times10^{11}$&$3.5\times10^{11}$\\
\end{tabular}
\caption{Characteristics of H$_3^+$ aurorae for the three VIMS long exposure time images.}
\label{tab_H3p_power}
\end{table*}

\begin{acknowledgments}
We thank D. Shemansky and H. M\'enager for providing synthetic H$_2$ spectra, the Community Coordinated Modeling Center for providing simulations results (ENLIL model), A. Sevellec for processing the data of the Kronos/RPWS database http://www.lesia.obspm.fr/kronos in Meudon, and engineers of Cassini/UVIS, VIMS and MIMI-INCA teams. LL thanks C. Tao, P. Schippers, U. Dyudina, B. Cecconi and X. Bonnin for useful discussions. French co-authors were supported by the CNES agency. PB and DM were supported in part by the NASA Office of Space Science under Task Order 003 of contract NAS5-97271 between NASA Goddard Space flight Center and the Johns Hopkins University, and by NASA NASA Grant NNX12AG81G.
\end{acknowledgments}

\end{article}
\end{document}